# Raman Signatures of Lithium Ion Dynamics in LLZO Garnet Electrolytes: Atomistic Insights from MD-Raman Calculations

*Takeru Miyagawa[1], Willis O'Leary[2], Manuel Grumet[1], Hyunwon Chu[2], Jennifer L. M. Rupp[3,4], Waldemar Kaiser[1*], and David A. Egger[1,5*]*

[1] Physics Department, TUM School of Natural Sciences, Technical University of Munich, 85748 Garching, Germany

[2] Department of Materials Science and Engineering, Massachusetts Institute of Technology, Cambridge, Massachusetts 02139, United States

[3] Fritz-Haber-Institute of the Max-Planck-Society, 14195 Berlin, Germany

[4] Department of Chemistry, TUM School of Natural Sciences, Technical University of Munich, 85748 Garching, Germany

[5] Atomistic Modeling Center, Munich Data Science Institute, Technical University of Munich, 85748 Garching, Germany

**Corresponding Authors**

Waldemar Kaiser, email: waldemar.kaiser@tum.de ;

David A. Egger, email: david.egger@tum.de

## ABSTRACT

Lithium lanthanum zirconate (LLZO) garnets are among the most promising solid electrolytes for next-generation batteries owing to their high ionic conductivity, chemical stability, and compatibility with lithium metal. Raman spectroscopy is commonly employed to distinguish the highly conductive cubic phase from the poorly conductive tetragonal phase of LLZO, yet the atomistic origin of these spectral differences and their direct connection to Li-ion transport remain unresolved. Here, we close this gap by comparing computed and experimental Raman spectra for the tetragonal, cubic, and Ta-doped variants of LLZO, with the computed spectra obtained from the MD-Raman approach that combines machine-learning molecular dynamics with first-principles polarizability calculations. We show that the contrasting ionic transport behavior across these LLZO variants is encoded in the vibrational dynamics of the lithium sublattice and gives rise to distinct features in their Raman spectra. A symmetry-resolved analysis further reveals that experimentally observed Raman peaks do not correspond to individual normal modes, but instead arise from overlapping contributions of multiple symmetry channels, challenging conventional peak-assignment approaches. By explicitly connecting experimentally accessible Raman signatures to the underlying atomic-scale dynamics, our results show how Raman spectroscopy can move beyond empirical phase identification toward a microscopic probe of Li-ion dynamics in lithium garnet electrolytes.

## Introduction

Lithium lanthanum zirconate ($Li_7La_3Zr_2O_{12}$, LLZO) garnets are among the most promising solid electrolytes for next-generation all-solid-state batteries due to their wide electrochemical stability window, high ionic conductivity in the cubic phase, and compatibility with lithium metal.[1–3] However, formation of the desirable fast conducting cubic phase can be a challenge.[4] In its tetragonal phase, LLZO exhibits too low Li-ion conductivity to serve as a battery electrolyte. When synthesized at elevated temperatures and with the introduction of aliovalent dopants, LLZO crystallizes in the conductive cubic phase with around 1 mS/cm at ambient conditions.[5–9] Although the associated symmetry change induces only minor changes to the host framework and lattice parameters, it fundamentally reorganizes the Li sublattice by altering the tetrahedral and octahedral site occupancies and thereby unlocks fast Li ion diffusion.[10] The transport characteristics of LLZO thus depend drastically on the balance between phase stabilization and lithiation level.[11] This balance is difficult to control in practice: LLZO frequently forms tetragonal/cubic phase mixtures rather than a single well-defined phase, and its synthesis, whether as powder, pellet, or thin film, is inherently prone to lithium loss, which destabilizes the cubic phase.[12,13] The ability to efficiently identify, understand, and control these phases is therefore a prerequisite for LLZO's practical use in next-generation cell architectures.

Among available experimental techniques, Raman spectroscopy serves as a powerful diagnostic tool for phase identification.[1,14] It offers rapid, nondestructive measurements and is highly sensitive to structural distortions. This has enabled the differentiation between cubic, tetragonal, and amorphous LLZO in both bulk and thin-film samples.[13,15,16] Earlier studies by Tietz et al.[14] and Mukhopadhyay et al.[17] proposed symmetry assignments for several Raman-active modes, which have since been widely used to assess phase purity[15,18–21] and to monitor dopant incorporation.[22,23] However, these assignments are inherently based on crystallographic symmetries and averaged structural motifs. As a result, they provide limited insight into how the complex dynamics, particularly Li-ion diffusion, shape the Raman response. This limitation becomes particularly relevant given the structural and dynamical complexity of LLZO. $Li^+$ occupies both tetrahedral and octahedral sites within the garnet framework, forming distinct sublattices with different coordination environments,[24] connected by a network of conduction pathways that vary across different phases, temperatures, and dopant chemistries.[25,26] Because Li-site environments and conduction pathways vary among LLZO phases, we hypothesize that the associated atomic motions produce distinct Raman-active vibrational signatures that can reliably differentiate highly ion-conducting from poorly conducting phases.

First-principles Raman calculations can, in principle, resolve this hypothesis by providing a direct link between Raman peaks and atomic-scale dynamics. Conventional approaches evaluate the derivative of the polarizability tensor along harmonic phonon eigenmodes, but this description becomes inadequate in the presence of strong anharmonicity and disorder. The harmonic normal-mode picture itself relies on a well-defined reference structure at a potential-energy minimum, which ceases to be meaningful when ions diffuse and their local environments are continually exchanged, as emphasized for liquids and aqueous systems where mode-based decompositions break down.[27] Such effects are, in fact, intrinsic to solid electrolytes and closely tied to their ionic conductivity. Anharmonic host-lattice dynamics are associated with pronounced low-energy vibrations that have been discussed as a driving force for ion conduction.[28–31] Ion migration itself is likewise an anharmonic phenomenon[32] and can leave a direct spectroscopic fingerprint; in liquid-like ion conductors it produces Raman central peaks,[29,33] mirrored by analogous soft-mode activity in quasielastic neutron scattering.[34–36] Capturing these signatures therefore requires going beyond the harmonic approximation. This can be achieved by molecular dynamics (MD)-based Raman approaches, in which the spectrum is obtained from the time-correlation function of the polarizability sampled along an MD trajectory rather than from harmonic phonon eigenmodes.[37–40] By construction, such approaches account for thermal motion, anharmonicity, and ion dynamics on equal footing, making them well suited to disordered and fast-ion-conducting systems.[33,40]

Here, we establish the atomistic link between Li ion migration and the Raman features in LLZO. To do so, we compute the Raman spectra for tetragonal (t-LLZO), cubic (c-LLZO), and tantalum-doped (Ta-LLZO) forms of LLZO using the MD-Raman approach, combining machine-learning molecular dynamics (MLMD)[41–43] with polarizabilities obtained from density functional perturbation theory (DFPT) along the resulting MLMD trajectories. With this method, we track how changes in the Li sublattice manifest in the Raman response and resolve the symmetry character of the contributing vibrational modes. The connection between Raman spectra and Li-ion dynamics reveals clear relationships between the motion of Li ions in the different sublattices and Raman signatures that differentiate conductive from non-conductive LLZO phases. Our results establish Raman spectroscopy as an interpretable probe of Li-ion dynamics in LLZO and highlight the potential of MD-Raman approaches to enable predictive relationships between Raman signatures and ion-transport across broader classes of solid electrolytes.

## Results and Discussion

We begin by evaluating the dynamical properties of the different LLZO phases. To this end, MLMD simulations were performed using force fields trained on-the-fly[41,42] against density functional theory (DFT) reference energies and forces for each respective LLZO phase (see Methods and Materials section). Quantitatively, the MLFFs achieve root-mean-square errors (RMSEs) in the range of ~23–95 meV $Å^{-1}$ across all chemical elements and LLZO phases, with Li showing the smallest deviations (~27.5 meV $Å^{-1}$) and heavier cations such as Zr exhibiting larger errors (~85 meV $Å^{-1}$), as visible in the parity plots provided in Figures S1–S3. Importantly, a comparable level of accuracy is maintained for c-LLZO at elevated temperature (900 K) and for Ta-LLZO, where Ta-containing environments are captured with similar fidelity. To validate these MLFFs specifically for the structural and vibrational properties relevant to this work, we further compare the radial distribution and vibrational density of states (VDOS) obtained from MLMD and reference AIMD simulations (Figure S4). The MLMD results closely reproduce both the local structural correlations and the VDOS obtained from AIMD, demonstrating hereby that the MLFFs accurately capture the quantities relevant to our vibrational and Raman analysis in this work. These results are further consistent with our previous MLFF benchmark,[43] where force errors of this magnitude were shown to accurately predict the ion transport mechanism and the VDOS at DFT accuracy in different classes of solid electrolytes. We therefore expect the same level of accuracy for MD simulations and subsequent vibrational analysis presented here.

MLMD simulations reveal distinct Li-ion transport behavior across the different phases. While t-LLZO exhibits no observable Li-ion diffusion on the simulated timescales of 200 ps, both c-LLZO and Ta-LLZO show significant ionic conduction, with estimated activation energies of 0.30 eV and 0.17 eV, respectively (Figure S5). The value obtained for c-LLZO agrees closely with previous simulations, which report activation energies of 0.30 eV[44] and 0.34 eV[2] for the cubic phase. Van Hove correlation analysis further distinguishes the transport mechanisms: t-LLZO shows no signatures of Li migration (Figure S6), c-LLZO exhibits concerted Li-ion motion (Figure S7), and Ta-LLZO is characterized by a $Li^+$ hopping-dominated diffusion mechanism (Figure S8). For the concerted mechanism, we identify that c-LLZO reproduces the picture established in earlier studies, in which tetrahedral- and octahedral-site Li ions migrate cooperatively rather than through isolated single-ion hops.[25,45]

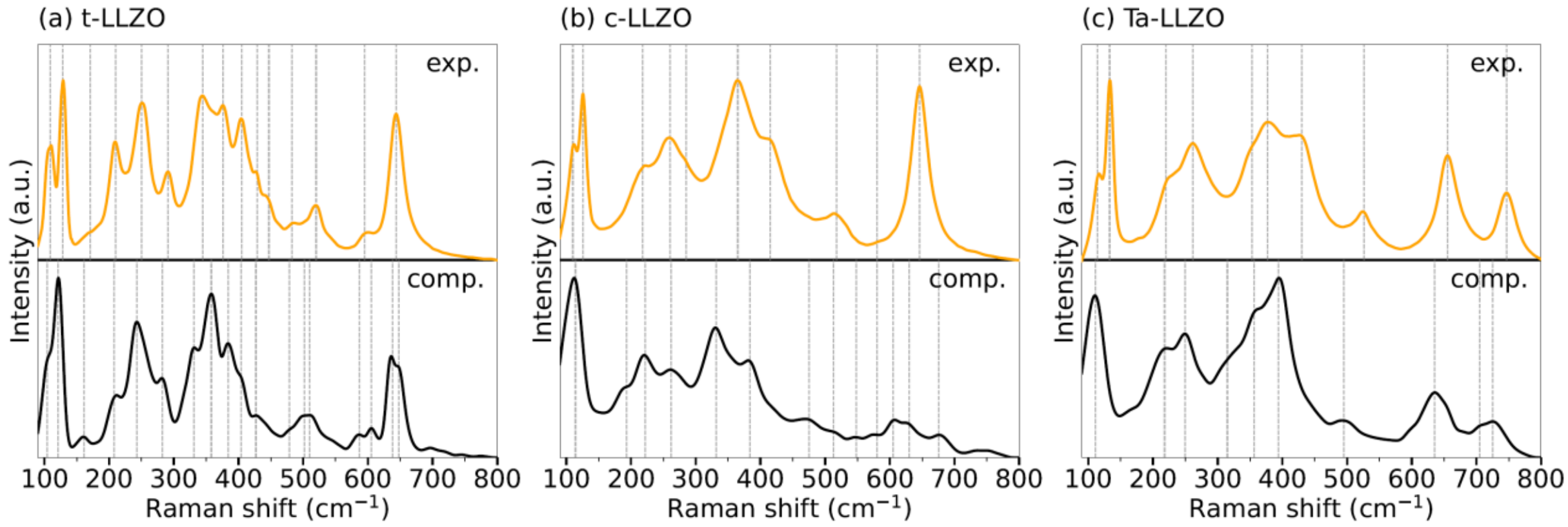

**Figure 1.** Finite-temperature Raman spectra of the undoped (a) tetragonal (t-LLZO) and (b) cubic LLZO (c-LLZO), and for the cubic (c) Ta-doped LLZO (Ta-LLZO) with polarizabilities from DFPT calculations. Black curves visualize the computed Raman spectra (comp.), while orange curves show experimental Raman measurements (exp.). Vertical lines highlight the peak position in the computed and measured Raman spectra.

Having established the distinct Li-ion transport mechanisms across the three phases, we now examine how these differences in the Li-ion dynamics manifest in the Raman response, comparing computed and experimental spectra for all three phases in Figure 1. The computed Raman spectrum of t-LLZO (Figure 1a) reproduces the peak positions and relative intensities of the experimental spectrum well. The spectrum exhibits a sequence of well-resolved peaks between 100 and 650 cm$^{-1}$, characteristic of a rigid host lattice with localized Li vibrations and no Li-ion diffusion. In the following, we discuss the different spectral regions in detail. A quantitative comparison of the peak positions obtained from theory and experiment is provided in Table S1-S3. Table S4 summarizes the experimental and computational LLZO compositions, Raman measurement conditions, and Li-ion transport properties.

The Raman spectra of c-LLZO and Ta-LLZO (Figures 1b and 1c), where Li ions are highly mobile (see Figure S5), differ substantially. The Raman spectrum of c-LLZO is characterized by broad, merged Raman features, consistent with the dynamic disorder in its highly mobile Li sublattice. Both experimental and computed spectra display three characteristic regimes: a low-frequency band near 150 cm$^{-1}$, a broad mid-frequency structure between 250 and 500 cm$^{-1}$, and a high-frequency band near 640 cm$^{-1}$. The high-frequency $ZrO_6$ stretching mode near 640 cm$^{-1}$ is present in both phases, reflecting the structural rigidity of the Zr–O framework regardless of Li-ion dynamics. In contrast, the mid-frequency region between 250 and 500 cm$^{-1}$ changes considerably. The sharp, well-resolved peaks of t-LLZO transition into a broad continuum in c-LLZO. This broadening is consistent with the rapidly fluctuating Li–O coordination environments as Li ions jump between the Li sites, consistent with the diffusive Li ion dynamics observed in the MSD and van Hove analyses (see Figures S6-S8). In the low-frequency region, the peak at 110 $cm^{-1}$ broadens and the peak at 160 $cm^{-1}$ disappears in c-LLZO. Both c-LLZO and Ta-LLZO additionally show enhanced Raman intensity around 180 $cm^{-1}$ relative to t-LLZO. Overall, the strongest broadening is observed in the modes involving Li motion and the associated Li–O Raman active modes, while host-lattice contributions remain sharp.

The Raman spectrum of Ta-LLZO closely resembles that of c-LLZO, consistent with their shared cubic phase and Li-ion conductivity. Subtle differences are nonetheless apparent. The peak at 391 $cm^{-1}$ is enhanced relative to the peak at 355 $cm^{-1}$ in both experimental and computed spectra of Ta-LLZO. The broadened Raman peaks in the Li-dominated frequency regimes persist in Ta-LLZO, confirming that diffusive Li dynamics are retained despite the lower simulation temperature. Most distinctively, Ta-LLZO exhibits an additional intensity in the high-frequency region near 730 $cm^{-1}$, which we assign to O–Ta–O vibrational modes involving oxygen atoms in the $TaO_6$ polyhedra, as evidenced by the projected VDOS shown in Figure S9. This band is absent in undoped LLZO and provides a direct spectroscopic marker of Ta incorporation. Experimentally, its intensity scales with Ta content, making it a quantitative probe of doping level and effectiveness.[8]

Given the pronounced disorder in the Li sublattice of c-LLZO and Ta-LLZO, one might expect their Raman response to resemble that of other highly disordered ion conductors. An important characteristic in the Raman response of some solid electrolytes, in which cations move within a strongly disordered sublattice, is the Raman central peak.[29,33] In systems where this feature occurs, the cations were reported to diffuse in a liquid-like manner. This causes the breakdown of Raman selection rules and renders low-frequency vibrations Raman active. [29,33] Interestingly, while the Li-sublattice in c-LLZO is also disordered and highly mobile, Raman central peaks are absent in both computed and experimental Raman spectra. This absence is consistent with $Li^+$ diffusion occurring within a crystallographically defined network of Li sites that preserve the overall lattice symmetry and with a comparatively limited relaxation of Raman selection rules, in contrast to systems such as α-AgI where liquid-like $Ag^+$ diffusion strongly disrupts the symmetry.[46] In contrast, the Raman selection rules for c-LLZO remain largely intact, and low-frequency Li vibrations remain weak in the Raman response.

To assign the Raman features described above to specific atomic motions, we now turn to the vibrational density of states (VDOS), shown in Figure 2. Across all LLZO systems, the low-frequency region below 200 $cm^{-1}$ is dominated by La vibrations, as expected given its large atomic mass. Zr ions contribute less prominently, with a main peak around 180 $cm^{-1}$ and weaker contributions extending up to 400 cm$^{-1}$. The vibrational signatures of both La and Zr remain largely unchanged across the different phases. In Ta-LLZO, Ta vibrations overlap with those of La without significantly modifying the La or Zr contributions. At higher frequencies above 200 cm$^{-1}$, the spectrum is dominated by vibrations of the lighter O and Li atoms. Comparing the LLZO phases, the oxygen VDOS shows subtle changes above 600 cm$^{-1}$. In c-LLZO, the O-related peaks broaden, and the distinct feature at 640 cm$^{-1}$ present in t-LLZO disappears. This high-frequency feature corresponds to stretching vibrations of O atoms coordinated to Zr.[14,17] Its disappearance in c-LLZO is consistent with a more disordered, less well-defined Zr–O coordination environment in the cubic phase. The vibrational contributions of oxygen in Ta-LLZO are intermediate between t-LLZO and c-LLZO, with an additional contribution emerges near 750 cm$^{-1}$ attributed to Ta–O bond vibrations.[47]

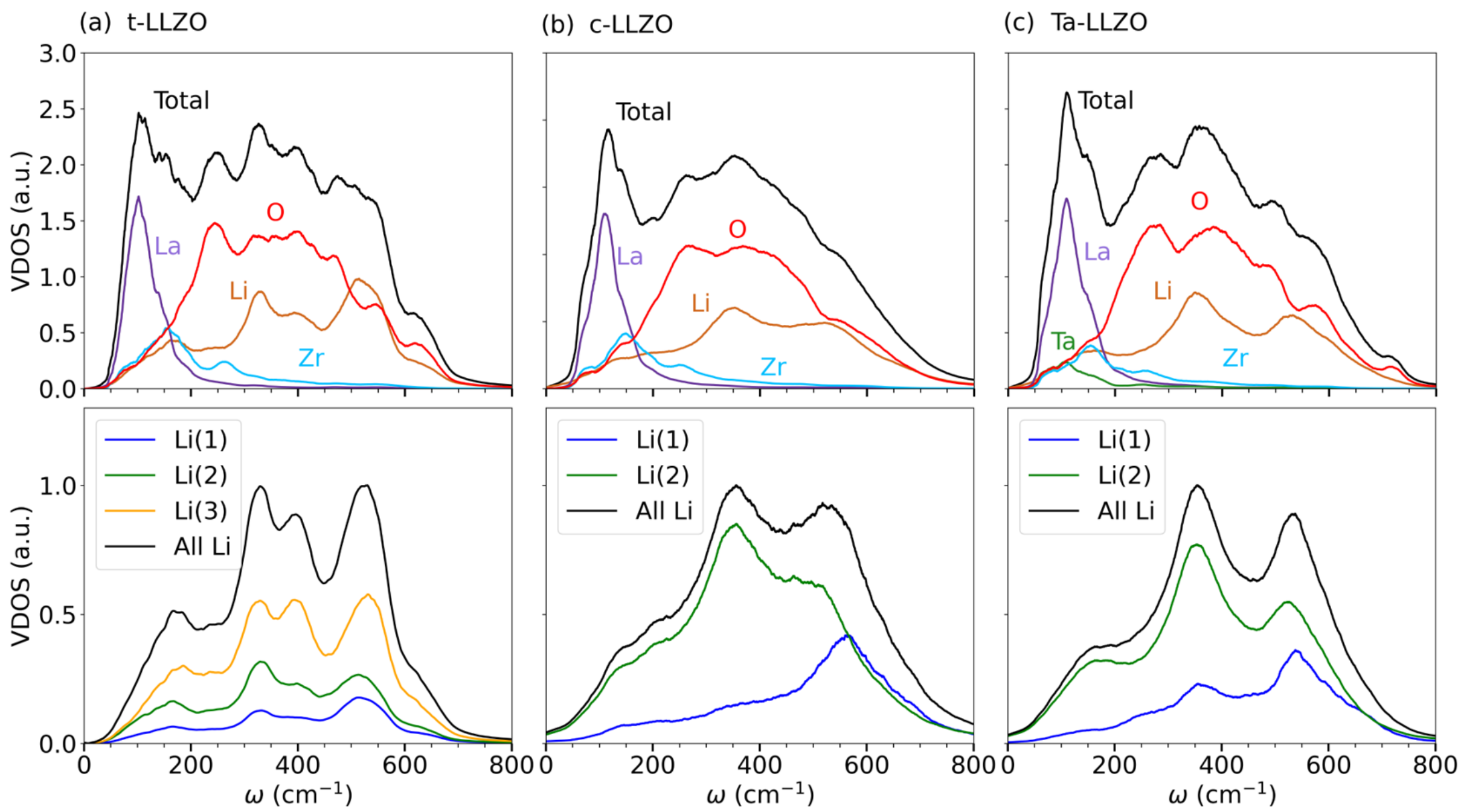

**Figure 2.** Vibrational density of states (VDOS) of (a) t-LLZO, (b) c-LLZO, and (c) Ta-LLZO. Top panels show the total VDOS (black) together with element-resolved contributions: Li (orange), La (purple), Zr (blue), O (red), and Ta (green, where applicable). Bottom panels resolve the Li contributions by crystallographic site and the total VDOS of all Li ions (All Li) with three distinct Li sites [Li(1), Li(2), Li(3)] for t-LLZO and with two Li sites each for c-LLZO and Ta-LLZO.

The most pronounced differences in the VDOS are observed in the Li sublattice. In t-LLZO, Li vibrations appear as distinct peaks at approximately 370 cm$^{-1}$, 420 cm$^{-1}$ and 560 cm$^{-1}$. These distinct peaks arise from well-defined Li–O coordination environments of t-LLZO, which exhibit narrow bond-length distributions (Figure S10). In contrast, the Li-projected VDOS of c-LLZO broadens significantly and largely loses its distinct peak structure. This broadening is consistent with the dynamic disorder in the cubic phase, where Li–O bond lengths are broadly distributed (see Figure S11) and Li ions are highly mobile, as reflected in the low activation energy for Li-ion diffusion (see Figure S5).[48] In particular, the high-frequency Li contribution around 560 $cm^{-1}$ is strongly suppressed and the 420 $cm^{-1}$ peak vanishes. The Li-projected VDOS of Ta-LLZO closely resembles that of c-LLZO, with the 420 $cm^{-1}$ peak likewise absent. The peaks at 370 and 560 $cm^{-1}$ are slightly more pronounced than in c-LLZO, which we attribute to the lower simulation temperature of Ta-LLZO.

To resolve the microscopic origin of these changes, we project the VDOS onto the crystallographically distinct lithium sites (Figure 2, lower panel). In t-LLZO, lithium fully occupies three ordered sites: the tetrahedral Li(1) (8*a*) site and the two octahedral Li(2) (16*f*) and Li(3) (32*g*). and the octahedral Li(2) and Li(3).[24,49] In the cubic phases c-LLZO and Ta-LLZO, lithium is instead distributed over the tetrahedral Li(1) (24*d*) and the octahedral Li(2) (96*h*) sites.[24,49,50] The two sets of sites are directly related. On going from the tetragonal to the cubic phase, the tetrahedral Li(1) (8*a*) maps to the cubic Li(1) (24*d*), while the two octahedral sites of t-LLZO, Li(2) (16*f*) and Li(3) (32*g*), merge into the partially occupied octahedral Li(2) (96*h*), over which Li is disordered and mobile.

These structural differences are directly reflected in the site-projected VDOS. In t-LLZO, the three Li sites contribute distinctly across the spectrum. Li(3) dominates the overall Li VDOS, reproducing its characteristic three peaks between 300 and 550 $cm^{-1}$. Li(2) contributes at lower intensity in the low- to mid-frequency range, while Li(1) is the weakest contribution overall with a modest peak concentrated at higher frequencies around 500 $cm^{-1}$. In c-LLZO, only two sites remain. Li(2) dominates the low- and mid-frequency range with a pronounced peak at 350-400 $cm^{-1}$, while Li(1) contributes mainly at higher frequencies with a peak near 550 $cm^{-1}$. Although both sites retain identifiable peaks, these are noticeably broader than the Raman peaks of t-LLZO, consistent with the dynamic disorder of the mobile Li sublattice in c-LLZO. The dominance of Li(2) at low frequencies is consistent with its central role in Li-ion diffusion,[10,51] while the persistent Li(1) feature at high frequencies reflects the more rigid tetrahedral $LiO_4$ environment (see Figure S11). This site-resolved picture explains the Raman response described above. The sharp mid-frequency peaks of t-LLZO arise from the well-defined Li(3) and Li(2) vibrations, and their broadening in c-LLZO, as these sites disorder and merge, produces the broad mid-frequency continuum observed in the c-LLZO Raman spectrum.

Ta-LLZO closely resembles c-LLZO in its site-projected VDOS. Li(2) again dominates the low- and mid-frequency range, with a maximum around 350-400 $cm^{-1}$, while Li(1) contributes at a lower intensity with but with discernible features in the same range. Compared to c-LLZO, the low-frequency shoulder at 180 $cm^{-1}$ is more pronounced and the Raman peaks are slightly sharper. Still, despite the lower temperature, the two-site character of the cubic phase is retained. Notably, in t-LLZO the vibrational mode at 420 $cm^{-1}$ originates mainly from the Li(3) site, with a partial contribution from Li(2). In c-LLZO and Ta-LLZO, the distinct Li(2) and Li(3) sites of t-LLZO merge into a single octahedrally coordinated site (Figure S11). The corresponding Li(2) vibration is blue-shifted and merges with the high-frequency feature near 550 $cm^{-1}$. This is directly reflected in the computed Raman spectra (Figure 1), where the resolved 420 $cm^{-1}$ feature of t-LLZO is no longer present as a distinct peak in c-LLZO or Ta-LLZO. The 420 $cm^{-1}$ peak thus offers a useful marker to identify the non-conductive tetragonal phase from the Raman response alone.

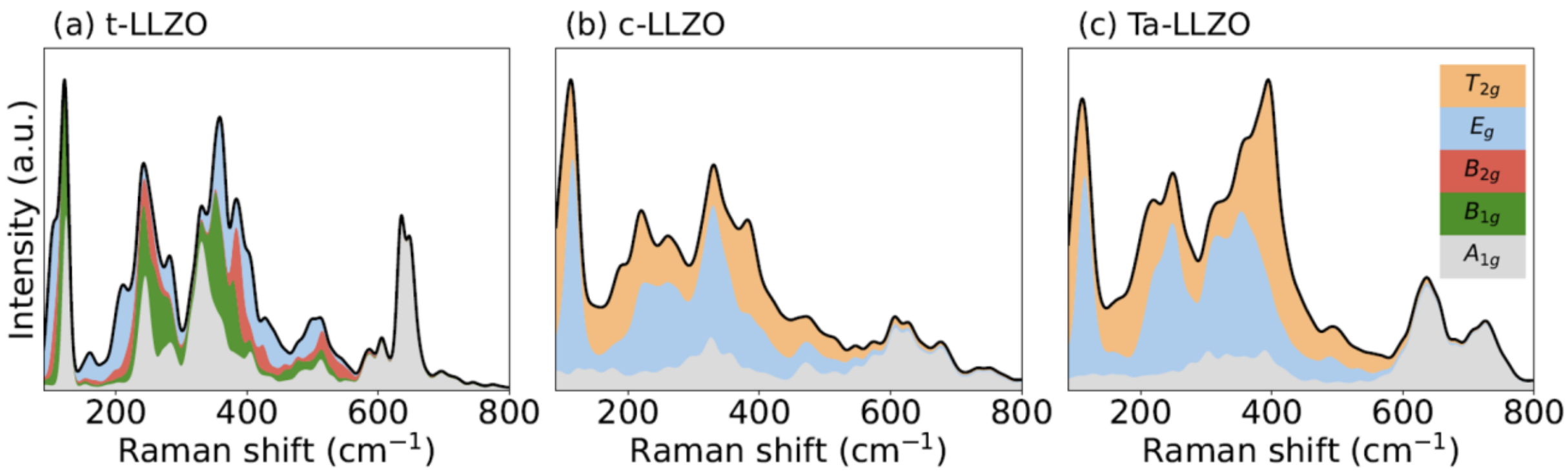

**Figure 3.** Symmetry-resolved Raman contributions for (a) t-LLZO, (b) c-LLZO, and (c) Ta-LLZO. Details of the decomposition in different symmetries can be found in the Supporting Information.

The site-resolved VDOS identifies which Li sites contribute to each spectral region, but not the symmetry of the underlying vibrations. To resolve this, we compute all components of the polarizability tensor from DFPT and decompose the Raman response into contributions from the irreducible representations of the relevant point groups (Figure 3 and Table S1-S3). For each Raman peak, this decomposition quantifies the percentage contribution of every symmetry channel (Table S1-S3), revealing that several symmetry modes contribute simultaneously rather than a single mode. This avoids the ambiguity of earlier assignments, which relied on tentative symmetry families inferred from powder measurements where polarized Raman cannot isolate individual channels.[14]

The symmetry decomposition directly explains the spectral changes identified above. The persistent $ZrO_6$ mode near 640 $cm^{-1}$ is of $A_{1g}$ symmetry in all three phases (Tables S1-S3). As a fully symmetric vibration of the rigid Zr–O framework, it appears largely insensitive to changes in the Li sublattice and therefore remains essentially unchanged across the transition. The sharp mid-frequency peaks of t-LLZO, by contrast, carry the lower-symmetry $B_{1g}$ and $B_{2g}$ channels, which appear around 350 and 380 $cm^{-1}$. These channels are entirely absent in c-LLZO and Ta-LLZO, and their disappearance coincides with the collapse of the resolved tetragonal peaks into the broad mid-frequency continuum of the conductive phases. The 420 $cm^{-1}$ feature, of $E_g$ character in t-LLZO, likewise has no counterpart in the cubic phases, where the intensity in this region transforms as $T_{2g}$ instead.

This reorganization of symmetry channels underlies the spectral changes. The four channels active in t-LLZO ($A_{1g}$, $B_{1g}$, $B_{2g}$, $E_g$) reduce to two in the cubic phases ($E_g$, $T_{2g}$), as the lower-symmetry channels permitted by the tetragonal distortion vanish in the average cubic symmetry. The lower-symmetry $B_{1g}$ and $B_{2g}$ channels are therefore responsible for the sharp, resolved peaks in the mid-frequency Raman spectrum of t-LLZO, and their loss in the cubic phases is what gives rise to the broad, featureless continuum. Throughout, individual Raman peaks carry contributions from more than one symmetry channel, so the measured peaks cannot be assigned to single symmetry species, as is commonly done in literature. For example, the c-LLZO feature near 365 $cm^{-1}$ combines $E_g$ (60%), $A_{1g}$ (22%), and $T_{2g}$ (19%) contributions (Table S2); assigning it to a single symmetry species would capture only part of its character. Ta-LLZO retains this cubic symmetry character, with the Ta dopant adding intensity at higher frequencies rather than reintroducing any tetragonal channels.

## Conclusion

Next generation batteries rely on solid-state electrolytes to enable operation with high-energy-density lithium and silicon anodes. One example is the Li-garnet solid electrolyte LLZO, which exists in several phases and is typically optimized toward the highly conducting cubic polymorph. Its structure is governed by a strong interplay between lithiation level and symmetry, giving rise to cubic, tetragonal, and even pyrochlore phases, with the conductive cubic phase further stabilized by extrinsic doping such as with tantalum. Raman spectroscopy is an excellent method for tracking the associated vibrational changes and the breaking and restoring of structural symmetry. In this work, we have investigated the atomistic origins of Raman signatures in lithium garnet electrolytes and their direct connection to Li-ion transport. To this end, we computed MD-Raman spectra for undoped tetragonal and cubic, as well as for Ta-doped LLZO phases by combining MLMD with polarizability calculations, an approach that captures the anharmonic vibrations and Li-ion diffusion inaccessible to harmonic Raman calculations.

The three phases of LLZO showed qualitatively distinct Li-ion dynamics that were reflected in corresponding changes in their Raman response. The ordered Li sublattice of t-LLZO yielded sharp, well-resolved peaks, whereas the highly mobile Li sublattices of c-LLZO and Ta-LLZO produced broad, merged features. Our analysis of the dynamical properties of individual lithium sites revealed the microscopic origin of these differences: well-defined coordination environments of the three distinct Li sites in t-LLZO give sharp signatures a ~370, ~420, and ~560 $cm^{-1}$, which broaden and reorganize as the octahedral sites merge and Li ions exchange rapidly in the cubic phases. The 420 $cm^{-1}$ feature is a distinctive marker of the non-conductive tetragonal phase, as it shifts upward and merges with the higher-frequency band in the cubic phases, where it is no longer resolved as a separate peak. Notably, despite the high Li-ion mobility, Raman central peaks are absent in both conducting phases, because $Li^+$ ions diffuse along well-defined pathways that do not cause a relaxation of all Raman selection rules.

By decomposing the Raman spectral signatures according to their symmetry, we found that the lower-symmetry $B_{1g}$ and $B_{2g}$ channels active in t-LLZO produce its sharp mid-frequency peaks. In the conducting phases, the higher average cubic symmetry switches these channels off, which leaves only the $E_g$ and $T_{2g}$ channels and gives rise to the broadening of Raman peaks. This reveals that the experimental Raman peaks reflect overlapping symmetry-allowed contributions rather than single symmetry channel, challenging recent assignments of individual peaks to single symmetry modes.[14,18] Such assignments are still useful as a shorthand and make results easier to communication, but they hide the fact that each peak comes from several modes across different symmetry channel.

Together, these results show how a strong connection between lithium sublattice dynamics and the Raman response of lithium garnet electrolytes, and how computed MD-Raman spectra can resolve the atomistic origin of features that experiment alone cannot disentangle. The framework is, in principle, transferable to other solid electrolytes, offering a route to connect Raman fingerprints to lithium coordination environments and ion dynamics. In addition, the DFPT calculations of the polarizability tensors, which dominates the computational cost of the MD-Raman approach, can be replaced by faster surrogate models, including machine-learned polarizability tensors,[33,40,52,53] Δ-machine-learning schemes,[54] bond-polarizability models,[55] and related approaches,[56,57] and with this predict Raman spectra at near-first-principles accuracy while substantially reducing cost. Overall, our work supports Raman spectroscopy as an interpretable diagnostic tool for the design of solid-state battery electrolytes.

## Methods and Materials

**Machine-Learning Molecular Dynamics.** Machine-learning force fields (MLFFs) were trained on-the-fly[41] against DFT energies and forces using the Vienna Ab initio Simulation Package (VASP),[58] with the projector augmented-wave method,[59] the PBE exchange–correlation functional,[60] Γ-point sampling, and a 1 fs time step in an NVT ensemble with the temperature being controlled with a Nosé-Hoover thermostat.[61,62] Separate MLFFs were trained for t-LLZO, c-LLZO, and Ta-LLZO; the training protocols and collected configuration counts are detailed in the Supporting Information. Relative to DFT, the force root-mean-square errors span 23–95 meV $Å^{-1}$ across all species and phases, with Li exhibiting the smallest deviation (27.5 meV $Å^{-1}$) and heavier cations such as Zr the largest (Figures S1–S3). Production trajectories of 100 ps, preceded by 20 ps of equilibration, were run at 300 K for t-LLZO and Ta-LLZO and at 900 K for c-LLZO.

**Structure Models.** The simulations employed unit cells of t-LLZO ($Li_7La_3Zr_2O_{12}$), c-LLZO, and Ta-LLZO ($Li_{6.375}La_3Zr_{1.375}Ta_{0.625}O_{12}$), with lattice parameters adopted from reported structures.[14,63,64] In Ta-LLZO, $Li^+$ vacancies were introduced to compensate the aliovalent substitution of $Zr^{4+}$ by $Ta^{5+}$. For the cubic phases, initial $Li^+$ configurations over the partially occupied sublattice were selected by an electrostatic-energy criterion evaluated through Ewald summation, iteratively removing the $Li^+$ ion whose removal most reduced the electrostatic energy until the target stoichiometry was reached. Complete cell parameters and atom counts are provided in the Supporting Information.

**Transport and Vibrational Analysis.** Li-ion diffusion coefficients were obtained from the mean-squared displacement via the Einstein relation. To account for thermal expansion, NPT simulations with a Langevin thermostat were performed at multiple temperatures, with four independent 200 ps runs per temperature, and activation energies were extracted from Arrhenius fits[65] of the temperature-dependent diffusion coefficients (Supporting Information, eqs S1–S3). Transport mechanisms were characterized using the van Hove[66] correlation function, and the vibrational density of states was computed for each atomic species and crystallographic Li site from the velocity autocorrelation function (eqs S4–S6).

**MD-Raman Spectra.** Raman spectra were computed from the time evolution of the polarizability tensor along the MLMD trajectories.[37,40] The polarizability was evaluated by density functional perturbation theory (DFPT) in VASP for every tenth trajectory snapshot (4000 DFPT calculations per system), and spectra were obtained within the Placzek approximation[67] from the polarizability time-correlation function (eq S7). To resolve the symmetry character of the Raman-active vibrations, the time derivative of the polarizability tensor was decomposed into one isotropic and five anisotropic symmetry-adapted channels and recombined according to the point group of each phase, $D_{4h}$ for t-LLZO and $O_h$ for the cubic phases.[68,69] Channel definitions and symmetry combinations are given in the Supporting Information (eqs S8–S14). Convergence analyses of MD-Raman calculations with respect to sampling (Figure S12) and trajectory length (Figures S13) are provided in the Supporting Information.

**Sample Synthesis.** t-LLZO was prepared by a sol–gel route,[20] c-LLZO by Ga-doping of the tetragonal precursor to a nominal composition of $Li_{6.4}Ga_{0.2}La_3Zr_2O_{12}$, and Ta-LLZO ($Li_{6.5}La_3Zr_{1.5}Ta_{0.5}O_{12}$) by solid-state reaction.[8,70] Full precursor quantities, calcination profiles, and sintering conditions are given in the Supporting Information.

**Raman Spectroscopy.** Raman spectra of the mechanically consolidated LLZO powders were recorded on a confocal WITec alpha300 R microscope using a 532 nm excitation laser at 10 mW, a 50× objective, and a 300 grooves $mm^{-1}$ grating, with 15 accumulations of 15 s each. Spatial reproducibility (Figure S14), influence of laser power and acquisition time (Figure S15), and Raman spectra for different Ta-concentations (Figure S16) are given in the Supporting Information.

## Declaration of Interests

The authors declare no competing interests.

## Data Availability Statement

The data underlying this study are openly available in Zenodo at DOI: 10.5281/zenodo.23526758.

## Supplemental Information

Computational details of machine-learning force field (MLFF) training and validation, molecular dynamics (MD) simulation and density functional theory (DFT) settings; description of the computed structural and dynamical properties, including mean-squared displacement, activation energies, and vibrational density of states; details of the MD-Raman calculations; full synthesis procedures and Raman measurement details; force parity plots for the MLFFs (Figures S1 to S3); MLMD accuracy analysis (Figure S4); and additional data comprising

Arrhenius plots (Figure S5), Van Hove correlation analysis (Figures S6 to S8), site-resolved VDOS (Figure S9) and bond-length analysis (Figures S10 and S11), and symmetry-resolved spectral analysis (PDF).

**Acknowledgement**

Funding provided by the Deutsche Forschungsgemeinschaft *via* Germany's Excellence Strategy – EXC 2089/2-390776260, and by the TUM-Oerlikon Advanced Manufacturing Institute, are gratefully acknowledged. The authors further acknowledge the Gauss Centre for Supercomputing e.V. for funding this project by providing computing time through the John von Neumann Institute for Computing on the GCS Supercomputer JUWELS at Jülich Supercomputing Centre.

**TOC Graphics**

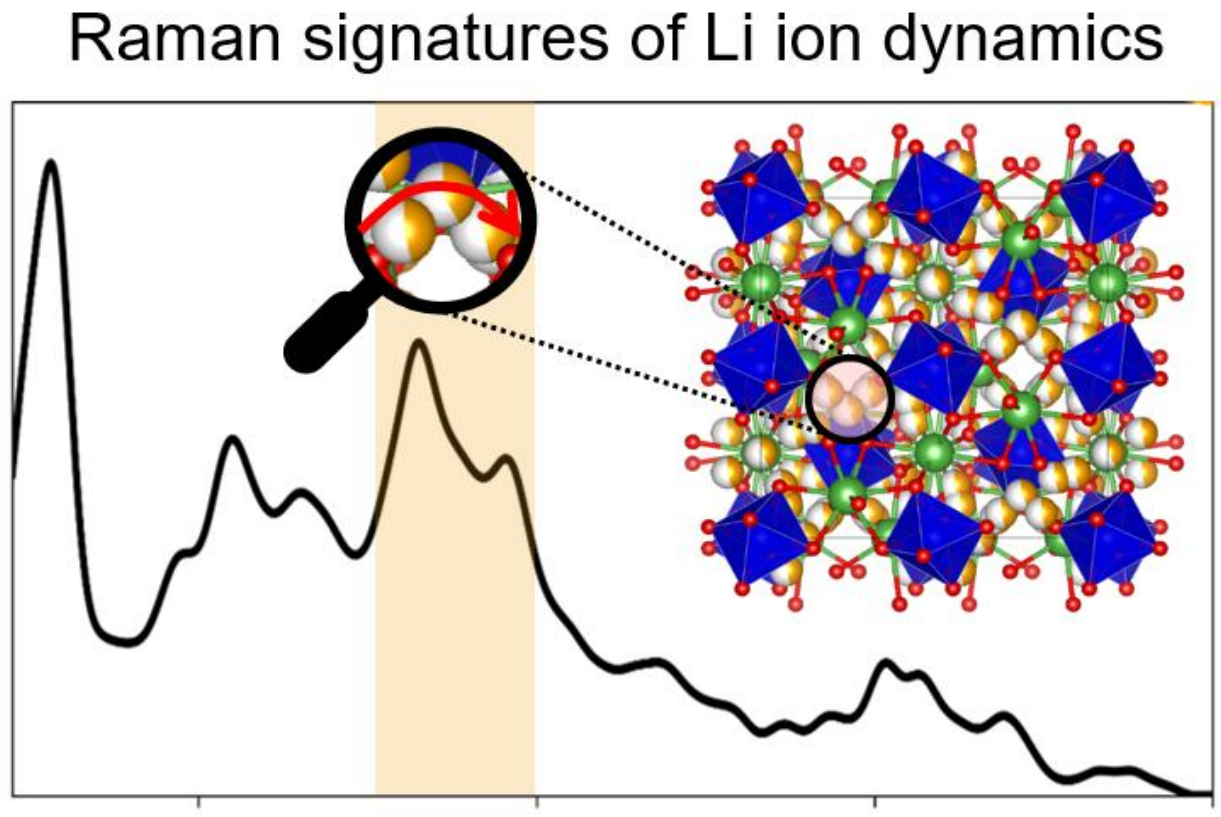

# *Supporting Information*

# Raman Signatures of Lithium Ion Dynamics in Li-Garnet Electrolytes: Atomistic Insights from MD-Raman Calculations

*Takeru Miyagawa[1], Willis O'Leary[2], Manuel Grumet[1], Hyunwon Chu[2], Jennifer L. M. Rupp[2,3], Waldemar Kaiser[1*], and David A. Egger[1,4,5*]*

[1] Department of Physics, TUM school of Natural Sciences, Technical University of Munich, 85748 Garching, Germany

[2] Department of Materials Science and Engineering, Massachusetts Institute of Technology, Cambridge, Massachusetts 02139-4307, United States

[3] Department of Chemistry, TUM school of Natural Sciences, Technical University of Munich, 85748 Garching, Germany

[4] Atomistic Modeling Center, Munich Data Science Institute, Technical University of Munich, Germany

[5] Munich Center for Machine Learning, Munich, Germany

**Corresponding Authors**

Waldemar Kaiser, email: waldemar.kaiser@tum.de ;

David A. Egger, email: david.egger@tum.de

## Computational Details

Machine-learning molecular dynamics (MLMD) simulations were performed using the Vienna Ab initio Simulation Package (VASP).[1,2] Machine-learning force fields (MLFFs)[3,4] were trained "on-the-fly" from DFT-calculated forces and energies along a molecular dynamics (MD) trajectory. The projector augmented wave (PAW) method[5,6] was employed along with the Perdew–Burke–Ernzerhof (PBE) exchange-correlation functional.[7] PAW potentials were used for Li (3 valence electrons: $1s^2$, $2s^2$), La (11 valence electrons: $4f^{0.0001}$, $5s^2$, $5p^6$, $5d^{0.9999}$, $6s^2$), Zr (12 valence electrons: $4s^2$, $4p^6$, $4d^3$, $5s^1$), O (6 valence electrons: $2s^2$, $2p^4$), and Ta (11 valence electrons: $5p^6$, $5d^4$, $6s^1$). The Brillouin zone was sampled exclusively at the Gamma point. All MLMD simulations were conducted using a time step of 1 fs. The number of Bessel functions was set to 8 for both the radial and the angular parts, and the cutoff radius for radial and angular descriptors was set to 5 Å.

We first trained the MLFFs for t-LLZO on-the-fly during MD simulations where we heated the system from 0 K to 300 K throughout 3 ps with a time step of 1 fs within an NVT ensemble, where the temperature was controlled with a Nosé-Hoover thermostat.[8,9] After heating up, the system was equilibrated at 300 K for 10 ps. After this training run, 738 local atomic configurations were collected for Li, 166 for La, 91 for Zr, and 962 for O atoms.

Starting with the t-LLZO MLFF, we continued the training for the c-LLZO structure by heating the temperature from 300 K to 1000 K for 3 ps. After the heating, the c-LLZO MLFF was trained at 1000 K for 32 ps. The extended training time for the high-temperature c-LLZO phase was chosen to sample a broader range of the configurational phase space, as the elevated temperature increases the probability of accessing diverse atomic environments and diffusion pathways.[10] 7454 local atomic configurations were collected for Li, 1587 for La, 1152 for Zr, and 7489 for O. For Ta-LLZO, we also starting the training procedure from the t-LLZO MLFF by heating up the temperature from 300 K to 400 K for 3 ps. After the heating, MLFF was trained on-the-fly for 17 ps. As a result, 3659 local atomic configurations were collected for Li, 957 for La, 537 for Zr, 466 for Ta, and 3657 for O.

After the training of MLFFs, we performed production runs for t-LLZO, c-LLZO, and Ta-LLZO using the aforementioned MLFFs. The temperature was set to 300 K for both t-LLZO and Ta-LLZO, and 900 K for c-LLZO. We equilibrated each system for 20 ps, followed by 100 ps production runs.

## System setup

We used a unit cell for t-LLZO, containing 56 Li, 24 La, 16 Zr, and 96 O with the cell parameters set as $a$=$b$=13.2361 Å, and $c$=12.7017 Å from the Materials Project ($Li_7La_3Zr_2O_{12}$, MP-942733).[11] For c-LLZO, the cell parameters were set as $a$=$b$=$c$=12.9827 Å,[12] containing 56 Li, 24 La, 16 Zr, and 96 O atoms. For tantalum-doped Ta-LLZO, we constructed a unit cell with the stoichiometry $Li_{6.375}La_3Zr_{1.375}Ta_{0.625}O_{12}$, with cell parameters $a$=$b$=$c$=12.9230 Å[13] and containing 51 Li, 24 La, 11 Zr, 5 Ta, and 96 O atoms. In Ta-LLZO, $Li^+$ vacancies were introduced to counterbalance the oxidation state change caused by replacing $Zr^{4+}$ ions with $Ta^{5+}$ ions.

To select the initial coordinates of $Li^+$ ions in c-LLZO and Ta-LLZO, we initially populated all available lithium sites and assigned charges to each ion ($Li^+$, $La^{3+}$, $Zr^{4+}$, and $O^{2-}$). For c-LLZO, this resulted in 120 Li initially occupied sites. We next computed the change in electrostatic energy upon removal of each $Li^+$ ion using the Ewald summation method.[14] The $Li^+$ ion whose removal caused the biggest decrease in electrostatic energy was subsequently removed. This process was repeated until the desired amount of $Li^+$ ions was reached. For Ta-LLZO, five Zr atoms were replaced by Ta using the same Ewald summation criterion, and the number of $Li^+$ ions was subsequently reduced from 120 to 51 to ensure charge neutrality within the system.

## Accuracy of the Machine-Learning Force Fields

Figures S1–S3 compare the atomic forces predicted by the MLFFs with those obtained from DFT calculations for t-LLZO, c-LLZO, and Ta-LLZO, respectively. For each compound, the comparison is performed element-wise by plotting the ML-predicted forces against the corresponding ab initio forces for the relevant ionic species. The root-mean-square error (RMSE) for each species is indicated in the respective panel.

In t-LLZO (Figure S1), the RMSE values remain low for all species, with Li showing the smallest deviations (27.5 meV $Å^{-1}$) and Zr the largest (85.3 meV $Å^{-1}$), consistent with the broader force distributions experienced by heavier cations. For c-LLZO (Figure S2), evaluated at 900 K, the force accuracy remains comparable despite increased ionic mobility. In Ta-LLZO (Figure S3), the inclusion of Ta does not compromise MLFF performance; all species, including Ta, exhibit RMSEs within 23–95 meV $Å^{-1}$. Across all systems, the strong agreement between MLFF and DFT forces confirms that the MLFFs accurately capture the interatomic interactions required for reliable MD simulations and subsequent Raman calculations.

## Structural and dynamical quantities

To construct the Arrhenius plots, *NPT* simulations with the Langevin thermostat were performed at multiple temperatures using the aforementioned MLFF to ensure thermal expansion was accounted for in the simulation. For c-LLZO, simulations were conducted at 700, 750, 800, 850, 900, 950, and 1000 K. For Ta-LLZO, the temperatures were 250, 275, 300, 325, 350, 375, and 400 K. At each temperature, a production run of 200 ps was carried out. To improve statistical reliability, four independent simulations were performed per temperature. Diffusion coefficients were computed according to the Einstein relation:

$$D = \frac{1}{6t_{\mathrm{sim}}} \mathrm{MSD}(t)\,, \tag{S1}$$

with the total simulation time $t_{\mathrm{sim}}$ and the mean-squared displacement (MSD) given by

$$\mathrm{MSD}(t) = \langle \left(\boldsymbol{r}_i(t) - \boldsymbol{r}_i(0)\right)^2 \rangle\,, \tag{S2}$$

where $\langle \cdots \rangle_i$ denotes the average all over Li ions $i$, and $r_i(t)$ gives the positions of Li ion $i$ at time $t$.

Using the aforementioned diffusion coefficient, we calculated the activation energy, $E_a$, using Arrhenius relation:[15,16]

$$D(T) = D_0 \exp\left(-\frac{E_a}{k_B T}\right), \tag{S3}$$

from which the activation energy $E_a$ was extracted from a linear fit of ln(*D*(*T*)) versus 1/*T*.

We further computed the vibrational density of states (VDOS) of an atomic species $i$ as the Fourier transform of the velocity autocorrelation:

$$\mathrm{VDOS_i}(\omega) \propto \frac{N_i m_i}{V} \mathcal{F}\{\langle \mathbf{v}_{i,j}(t+\tau) \cdot \mathbf{v}_{i,j}(\tau) \rangle_j\}, \tag{S4}$$

with $\mathbf{v}_{i,j}(t)$ being the velocity of the *j*-th atom of type *i* at time *t*, and $\langle \ldots \rangle_j$ denotes the average over all atoms *j*. $N_i$, $m_i$, and *V* denote the number of contributing atoms, the mass of the atoms, and the simulation volume, respectively.

Li-site resolved VDOS were obtained by dynamically assigning each Li ion to its nearest crystallographic Li site, as defined in the corresponding crystallographic structure. In t-LLZO, Li occupies three crystallographically distinct sites, the tetrahedral Li1 (8a) site, and the octahedral Li2 (16f) and Li3 (32g) sites. In the cubic phases, Li is distributed over the tetrahedral Li1 (24d) and octahedral Li2 (96h) sites. To isolate the vibrational properties associated with a given site, Li ions were assigned to a crystallographic site only if they remained there for at least 3 ps. This criterion reduces discontinuities arising from site-to-site jumps and transient oscillations between neighboring sites, allowing us to focus on the local vibrational dynamics. In c-LLZO, several partially occupied 96h (Li2) sites are separated by only 0.8 Å to another, enabling rapid Li ion exchange between neighboring sites. To reduce assignment ambiguities and avoid discontinuities, adjacent Li2 sites were grouped into a single effective Li2 site located at their geometric center. The population of Li ions contributing to the site-resolved VDOS is therefore updated dynamically throughout the trajectory.

Lastly, we computed the van Hove correlation,[17] a measure that quantifies spatiotemporal correlations between $Li^+$ ions. The distinctive part, $G_d$, measures the radial distribution of $Li^+$ ions (indexed with $j$) with respect to a reference $Li^+$ ion of the same type (indexed with $i$) as a function of a time interval $\Delta t$:

$$G_d(r, \Delta t) = \frac{1}{4\pi r^2 N} \langle \sum_{i=1...N} \sum_{j=1...N; j \neq i} \delta(r - |\mathbf{r}_i(t+\Delta t) - \mathbf{r}_j(t)|) \rangle_t, \quad \text{(S5)}$$

with the Dirac delta function $\delta$, $N$ is the number of $Li^+$ ions, and the time average $\langle ... \rangle_t$. In contrast, the self-part of the van Hove correlation, $G_s$, provides spatiotemporal correlations of the radial distribution of ion $i$ with itself:

$$G_s(r, \Delta t) = \frac{1}{4\pi r^2 N} \langle \sum_{i=1...N} \delta(r - |\mathbf{r}_i(t+\Delta t) - \mathbf{r}_i(t)|) \rangle_t . \quad \text{(S6)}$$

**Site-resolved Li-O bond length analysis**

The Li–O bond-length probability densities were computed from *NVT* molecular dynamics simulations with a total production time of 100 ps. Li ions were classified into Li1 (24d) and Li2 (96h) sites based on their Wyckoff positions in the fully relaxed reference structure. For each Li site in the reference structure, the corresponding coordination polyhedron was defined by identifying its nearest O neighbors (four O atoms for Li1 and six O atoms for Li2). These polyhedral templates were constructed from the reference structure and used consistently throughout the trajectory. During the trajectory analysis, Li ions at each time step were assigned to the nearest reference Li1 or Li2 site based on a distance criterion. A Li ion was considered to occupy a given site only if its distance from the corresponding reference position was smaller than 2.0 Å. This ensured a consistent mapping between instantaneous Li positions and their underlying crystallographic sites. For Li ions satisfying this site criterion, Li–O bond lengths were computed with respect to the O atoms defining the corresponding polyhedral template. Distances were evaluated using the minimum-image convention in fractional coordinates and converted to Cartesian distances using the reference lattice matrix. To ensure well-defined coordination environments, additional geometric constraints were applied: all Li–O distances were required to be smaller than 3.5 Å.

For each Li coordination polyhedron (fourfold for Li1 and sixfold for Li2), the Li–O distances were ordered by magnitude at each time step. Distances occupying the same position in this ordered sequence were accumulated over the production trajectory. Probability density functions (PDFs) were computed by discretizing the distance range from 1.0 Å to 3.5 Å into 200 equally spaced bins and normalizing the histograms.

**MD-Raman calculations**

To compute the Raman spectra, we performed DFPT calculations in VASP along snapshots extracted from the MLMD trajectories to obtain the time evolution of the polarizability tensor $\boldsymbol{\alpha}(t)$, which is interchangeable with the dielectric tensor for solids. DFPT calculations were conducted for every 10th snapshot along the 40 ps long MLMD trajectories, resulting in 4,000 DFPT calculations per system.

In all cases, the Raman spectra were then computed within the Placzek approximation[18] from the time series of the polarizability tensor $\alpha(t)$ using

$$I(\omega) \propto \frac{(\omega_{\mathrm{in}} - \omega)^4}{\omega} \frac{1}{1 - \exp\left(-\frac{\hbar\omega}{k_{\mathrm{B}}T}\right)} \frac{45\alpha_\tau^2 + 7\gamma_\tau^2}{45}, \tag{S7}$$

where $\omega$ is the Raman shift and $\omega_{\mathrm{in}}$ corresponds to a 514 nm excitation, and $k_{\mathrm{B}}$ is the Boltzmann constant. The second term on the right-hand side is the Bose–Einstein weighting factor, which accounts for the thermal population of the vibrational modes in the computation of the Raman spectra. The isotropic invariant is defined from the autocorrelation of the trace of the dielectric tensor,

$$\alpha_\tau^2 = \frac{1}{9}\int_{-\infty}^{\infty} \langle\left(\dot{\alpha}_{xx}(\tau) + \dot{\alpha}_{yy}(\tau) + \dot{\alpha}_{zz}(\tau)\right)\left(\dot{\alpha}_{xx}(\tau + t) + \dot{\alpha}_{yy}(\tau + t) + \dot{\alpha}_{zz}(\tau + t)\right)\rangle_\tau \exp(-i\omega t)\, dt\ , \tag{S8}$$

and the anisotropic invariant is defined as,

$$\begin{aligned}\gamma_\tau^2 = {} & 3\int_{-\infty}^{\infty} \left(\langle\dot{\alpha}_{xy}(\tau)\dot{\alpha}_{xy}(\tau + t)\rangle_\tau + \langle\dot{\alpha}_{yz}(\tau)\dot{\alpha}_{yz}(\tau + t)\rangle_\tau + \langle\dot{\alpha}_{zx}(\tau)\dot{\alpha}_{zx}(\tau + t)\rangle_\tau\right) \times \exp(-i\omega t)\, dt \\ & + \frac{1}{2}\int_{-\infty}^{\infty} \langle\left(\dot{\alpha}_{xx}(\tau) - \dot{\alpha}_{yy}(\tau)\right)\left(\dot{\alpha}_{xx}(\tau + t) - \dot{\alpha}_{yy}(\tau + t)\right)\rangle_\tau \exp(-i\omega t)\, dt \\ & + \frac{1}{2}\int_{-\infty}^{\infty} \langle\left(\dot{\alpha}_{yy}(\tau) - \dot{\alpha}_{zz}(\tau)\right)\left(\dot{\alpha}_{yy}(\tau + t) - \dot{\alpha}_{zz}(\tau + t)\right)\rangle_\tau \exp(-i\omega t)\, dt \\ & + \frac{1}{2}\int_{-\infty}^{\infty} \langle\left(\dot{\alpha}_{zz}(\tau) - \dot{\alpha}_{xx}(\tau)\right)\left(\dot{\alpha}_{zz}(\tau + t) - \dot{\alpha}_{xx}(\tau + t)\right)\rangle_\tau \exp(-i\omega t)\, dt\,.\end{aligned} \tag{S9}$$

The autocorrelation function used in Eqs. (S5)–(S6) is defined as

$$\langle \dot{\alpha}_{\alpha\beta}(\tau)\dot{\alpha}_{\alpha\beta}(\tau+t)\rangle_\tau = \int_{-\infty}^{\infty} \dot{\alpha}_{\alpha\beta}(\tau)\dot{\alpha}_{\alpha\beta}(t+\tau)d\tau. \tag{S10}$$

Gaussian broadening with a width of 8 $\mathrm{cm}^{-1}$ was applied. All spectra were normalized within the 90–800 $\mathrm{cm}^{-1}$ range for comparison.

To analyze the symmetries of the modes contributing to the Raman activity, we decompose the calculated polarizability tensor into symmetry-adapted components according to the point-group symmetry of each phase.[19–21] One isotropic and five anisotropic channels were constructed as a linear combination of the time derivative of the polarizability tensor:

$$s_0(t) = \frac{\dot{\alpha}_{xx}(t) + \dot{\alpha}_{yy}(t) + \dot{\alpha}_{zz}(t)}{3}, \tag{S11}$$

$$s_1(t) = \sqrt{\frac{3}{4}}\left[\dot{\alpha}_{xx}(t) - \dot{\alpha}_{yy}(t)\right],$$

$$s_2(t) = \sqrt{\frac{1}{4}}\left[\dot{\alpha}_{xx}(t) + \dot{\alpha}_{yy}(t) - 2\dot{\alpha}_{zz}(t)\right],$$

$$s_3(t) = \sqrt{3}\dot{\alpha}_{xy}(t),$$

$$s_4(t) = \sqrt{3}\dot{\alpha}_{xz}(t),$$

$$s_5(t) = \sqrt{3}\dot{\alpha}_{yz}(t).$$

For each channel, the spectral weight was obtained via Fourier transformation:

$$S_k(\omega) = |\mathcal{F}[s_k(t)]|^2, \qquad (k = 0, \ldots, 5) \tag{S12}$$

These spectral weights were subsequently combined according to the symmetry of the crystal structure.

t-LLZO belongs to the point group $D_{4h}$, for which the Raman-active irreducible representations are $A_{1g}, B_{1g}, B_{2g}, E_g$.The symmetry-resolved Raman intensities were constructed as

$$I_{A_{1g}}(\omega) \propto S_0(\omega) + S_2(\omega), \tag{S13}$$

$$I_{B_{1g}}(\omega) \propto S_1(\omega),$$

$$I_{B_{2g}}(\omega) \propto S_3(\omega),$$

$$I_{E_g}(\omega) \propto S_4(\omega) + S_5(\omega).$$

$A_{1g}$ contains contributions from both the isotropic trace component and one diagonal anisotropic combination, while the doubly degenerate $E_g$ representation arises from the $xz$ and $yz$ tensor components. c-LLZO and Ta-LLZO were treated within cubic symmetry (point group $O_h$). The Raman-active irreducible representations are $A_{1g}, E_g, T_{2g}$. The symmetry-resolved intensities were constructed as,

$$I_{A_{1g}}(\omega) \propto S_0(\omega)\,, \tag{S14}$$
$$I_{E_g}(\omega) \propto S_1(\omega) + S_2(\omega)\,,$$
$$I_{T_{2g}}(\omega) \propto S_3(\omega) + S_4(\omega) + S_5(\omega)\,.$$

**Convergence of MD-Raman spectra**

The convergence of Raman spectra was assessed with respect to both the temporal spacing of DFPT calculations and the trajectory length. Sampling intervals of 10-20 fs yield nearly identical spectra, whereas larger intervals introduce deviations at high frequency caused by the Nyquist limit (Figure S12). We further evaluated the convergence of Raman spectra with trajectory length, using an increased amount of MD snapshots for the DFPT calculations with equal temporal spacing of 10 fs. We observe that beyond 3000 snapshots no major changes in Raman spectra are observed, as shown in Figure S13.

**Synthesis of LLZO Powders**

The tetragonal phase of LLZO (t-$Li_7La_3Zr_2O_{12}$, t-LLZO) was synthesized via a sol-gel method.[22] Stoichiometric amounts of $LiNO_3$ (99.99%, Merck), $La(NO_3)_3{\cdot}6H_2O$ (99.9%, Thermo Scientific), and zirconium(IV) 2,4-pentanedionate (Thermo Scientific) were dissolved in a mixed solvent of deionized water and absolute ethanol. This specific solvent combination was utilized to effectively dissolve both the nitrate and pentanedionate precursors, forming a stable sol. The solution was subsequently heated at 90 °C to evaporate the solvent, resulting in a homogeneous gel powder. The gel precursor was then transferred to alumina crucibles with lids and fired in a muffle furnace under ambient air. To ensure high phase purity and complete crystallization, a heating profile of 650 °C for 15 h was applied with a heating rate of 5 °C $min^{-1}$. The cubic phase of LLZO (c-LLZO) was prepared by introducing Ga-doping into the parent tetragonal structure to achieve a nominal composition of $Li_{6.4}Ga_{0.2}La_3Zr_2O_{12}$. Stoichiometric amounts of $Ga_2O_3$ (99.99%, Thermo Scientific) powder and the pre-synthesized t-LLZO precursor were thoroughly mixed and homogenized using an agate mortar. The resulting powder mixture was then annealed in a muffle furnace at 650 °C for 10 h under ambient air using a heating rate of 5 °C $min^{-1}$. During this step, the sample was processed directly in powder form covered with the parent powder to mitigate potential lithium loss. The Ta-doped LLZO (Ta-LLZO) with a composition of $Li_{6.5}La_3Zr_{1.5}Ta_{0.5}O_{12}$ was synthesized through a solid-state reaction route.[23,24] Stoichiometric amounts of $La(OH)_3$ (99.95%, Thermo Scientific), $ZrO_2$ (99.7%, Thermo Scientific), and $Ta_2O_5$ (99.5%, Thermo Scientific) were mixed with a 40 wt% excess of LiOH (99.995%, Thermo Scientific) to compensate for lithium volatilization at higher temperatures. The precursors were homogenized via planetary milling (PQ-N04, Across International, 500 rpm, 1 h) in absolute ethanol using $ZrO_2$ media, and subsequently dried at 90 °C. The mixture was placed in MgO crucibles and subjected to a two-step calcination process under a constant flow of synthetic air in a tube furnace: a primary calcination at 750 °C for 10 h (heating rate of 5 °C $min^{-1}$), followed by an intermediate ball-milling step for 12 h at 500 rpm, and a secondary calcination at 750 °C for 5 h (heating rate of 10 °C $min^{-1}$). Finally, the calcinated powder was sintered directly in a MgO crucible under a pure oxygen flow (50 sccm) at 1100 °C for 5 h, employing heating and cooling rates of 10 °C $min^{-1}$.

**Powder Raman Measurement**

The experimental Raman features of the LLZO samples were investigated via a confocal WITec alpha300 R Raman Imaging Microscope (WITec, Oxford Instruments). For the measurement, the LLZO powders were consolidated into pellet-like compacts by mechanical pressing to provide a smooth, uniform surface for accurate peak position and intensity evaluations. A 532 nm laser line (spectral center: 2,050 $cm^{-1}$) operated at 10 mW was employed as the excitation source to eliminate sample heating effects. A Zeiss EC Epiplan 50× objective lens and a 300 grooves $mm^{-1}$ grating were used for signal collection and dispersion, respectively. Spectral acquisition involved 15 accumulations with an exposure time of 15 s per scan.

Raman spectra were recorded at two different locations on each LLZO sample. As shown in Figure S14, the spectra obtained from different positions exhibit comparable spectral profiles, with only minor variations in relative intensities and peak positions. The observed spectra are furthermore consistent with previous reports of t-LLZO, c-LLZO, and Ta-LLZO.[12,22,25]

To assess possible laser-induced effects during Raman measurements, additional spectra of Ta-LLZO were collected at laser powers of 5, 8, and 10 mW under otherwise identical acquisition conditions. An additional spectrum was recorded at 5 mW using a shorter acquisition time. As shown in Figure S15, the spectra exhibit nearly identical peak positions and overall profiles, with no detectable peak shifts or additional Raman bands. The shorter acquisition results only in an increased noise level. The absence of detectable spectral changes across the investigated laser powers supports the reliability of the Raman measurements performed at 10 mW.

**Comparison of Computational and Experimental Compositions**

Direct comparison between cubic LLZO calculations and experiments is complicated by the fact that stoichiometric LLZO adopts the tetragonal phase at room temperature, whereas stabilization of the highly conductive cubic phase requires doping. Ga-doped LLZO was therefore selected as the experimental cubic reference compound because Ga effectively stabilizes the cubic phase while introducing negligible additional Raman features compared to transition-metal dopants such as Ta or Nb.[25] The resulting Raman spectrum retains the characteristic broad features of cubic LLZO, particularly in the frequency range associated with Li dynamics. Based on these considerations, we selected $Li_{6.4}Ga_{0.2}La_3Zr_2O_{12}$, following the established composition and experimental characterization by some of us.[22]

For Ta-LLZO, the experimental composition $Li_{6.5}La_3Zr_{1.5}Ta_{0.5}O_{12}$ and the simulated composition $Li_{6.375}La_3Zr_{1.375}Ta_{0.625}O_{12}$ both lie within the compositional range known to stabilize the highly conductive cubic phase. To assess the influence of Ta content on the Raman response, we additionally compared experimental spectra across a range of Ta concentrations (Figure S16). At lower Ta concentrations, more distinguishable vibrational features appear in the mid-frequency region, whereas extensive substitution (x = 1.0) leads to pronounced changes in the Li-related spectral profile, suggesting modifications in the local Li environments. In contrast, the intermediate composition (x = 0.5) exhibits the broad Raman features characteristic of cubic LLZO, without the pronounced spectral differences observed at lower and higher Ta concentrations. The observed variations in the vibrational response agree with our previously reported influence of Ta substitution on Li-site environments.[24] The proximity of the calculated composition (x = 0.625) to the established x = 0.6 composition supports the selection of the computational model. The associated difference in Li vacancy concentration may influence quantitative transport properties, including Li diffusivity and activation energy. However, the principal Raman characteristics associated with the disordered cubic Li sublattice are expected to be preserved, supporting the interpretation of Li-ion dynamics using the selected computational composition.

In Table S4, we summarize all compositional and experimental LLZO compositions, as well as the experimental settings and key transport properties extracted from our simulations.

## Additional Figures

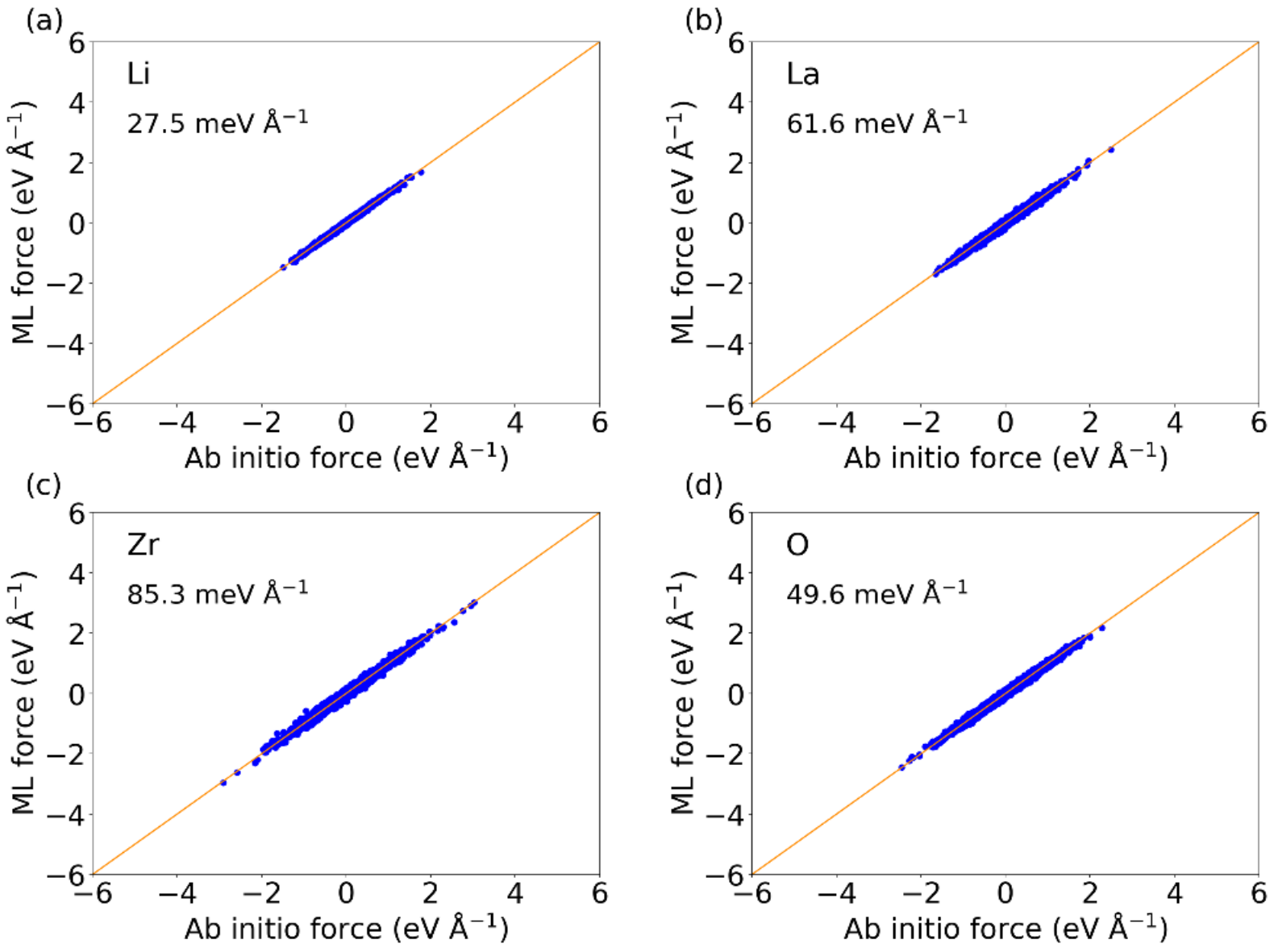

**Figure S1. Comparison of atomic forces calculated from the MLFF and DFT for t-LLZO.** Parity plots for (a) Li, (b) La, (c) Zr, and (d) O show excellent agreement between ML and DFT forces, with species-dependent RMSE values remaining low. A 50 ps MLMD simulation at 300 K was performed using a Nosé–Hoover thermostat, and 100 snapshots sampled every 0.5 ps were evaluated with both DFT and the MLFF.

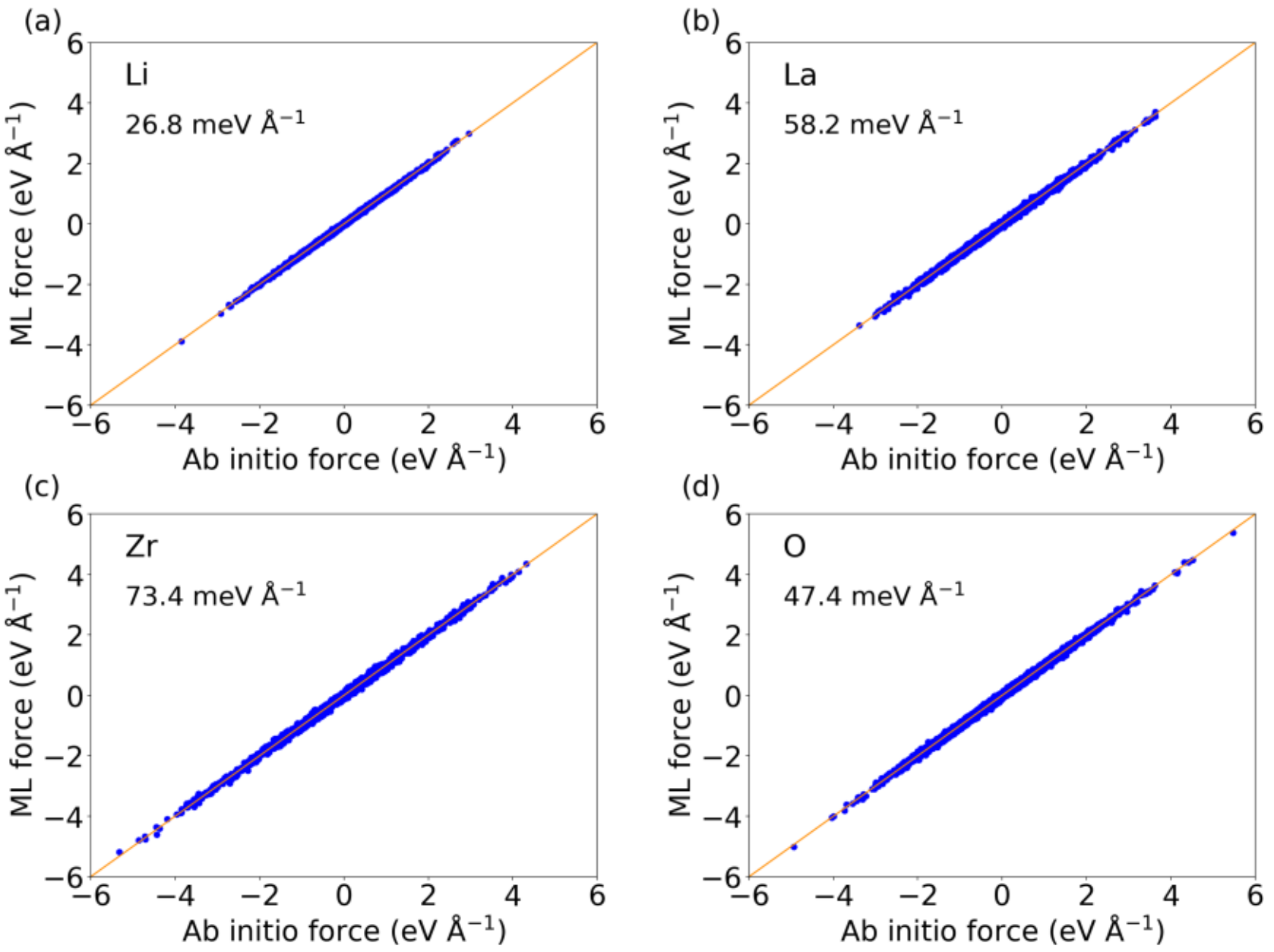

**Figure S2. Comparison of atomic forces calculated from the MLFF and DFT for c-LLZO.** Parity plots for (a) Li, (b) La, (c) Zr, and (d) O demonstrate similarly high accuracy at 900 K, characteristic of the cubic phase. The comparison uses 100 snapshots extracted every 0.5 ps from a 50 ps MLMD trajectory.

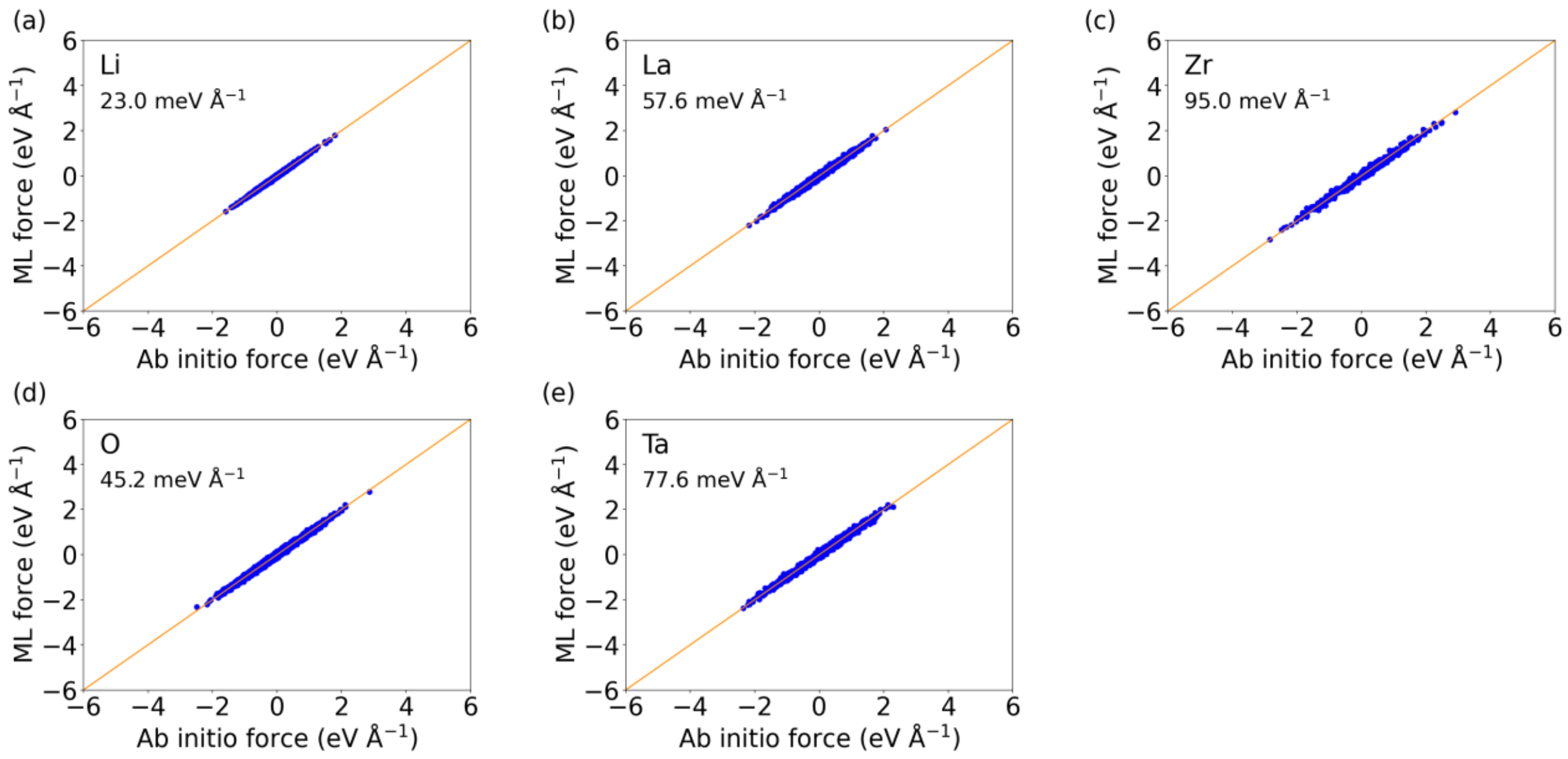

**Figure S3. Comparison of atomic forces calculated from the MLFF and DFT for Ta-LLZO.** Parity plots for (a) Li, (b) La, (c) Zr, (d) O, and (e) Ta show that the inclusion of Ta does not degrade MLFF performance. The dataset consists of 100 snapshots collected every 0.5 ps from a 50 ps MLMD simulation at 300 K.

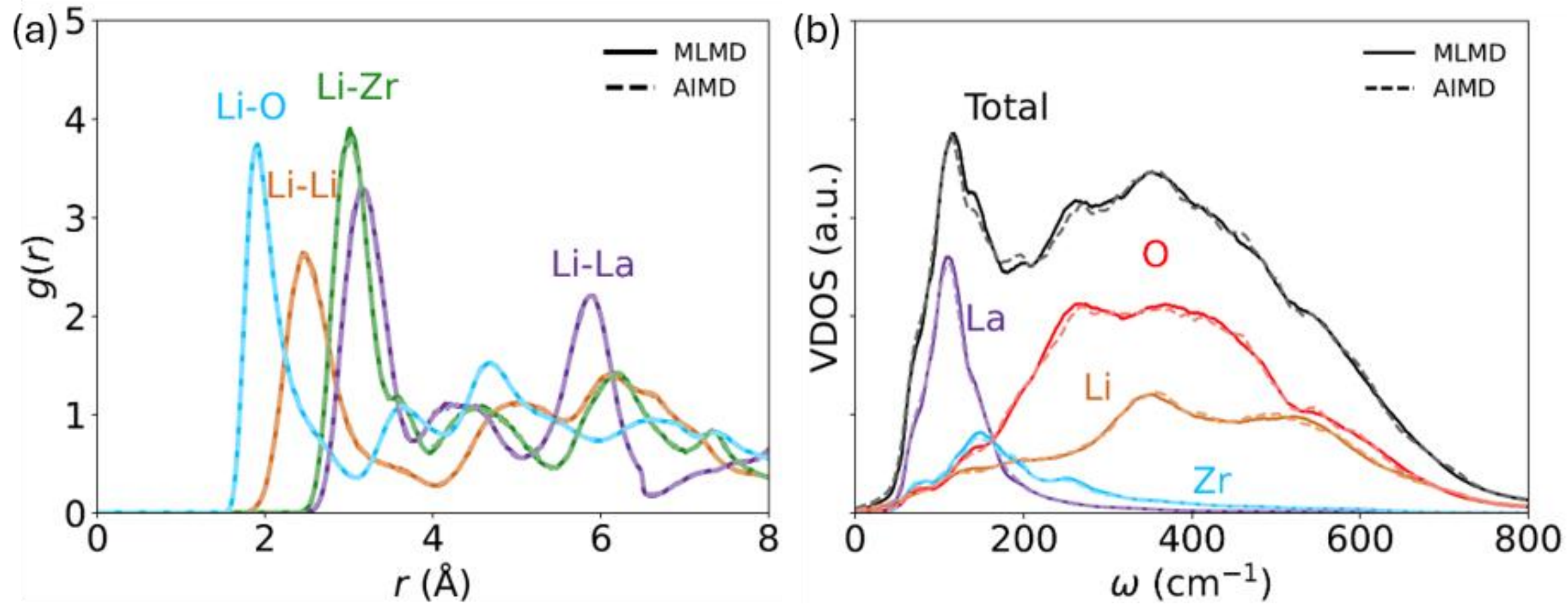

**Figure S4.** Comparison of (a) radial pair distribution and (b) vibrational density of states between AIMD and MLMD for c-LLZO, with a 10 ps AIMD reference simulation.

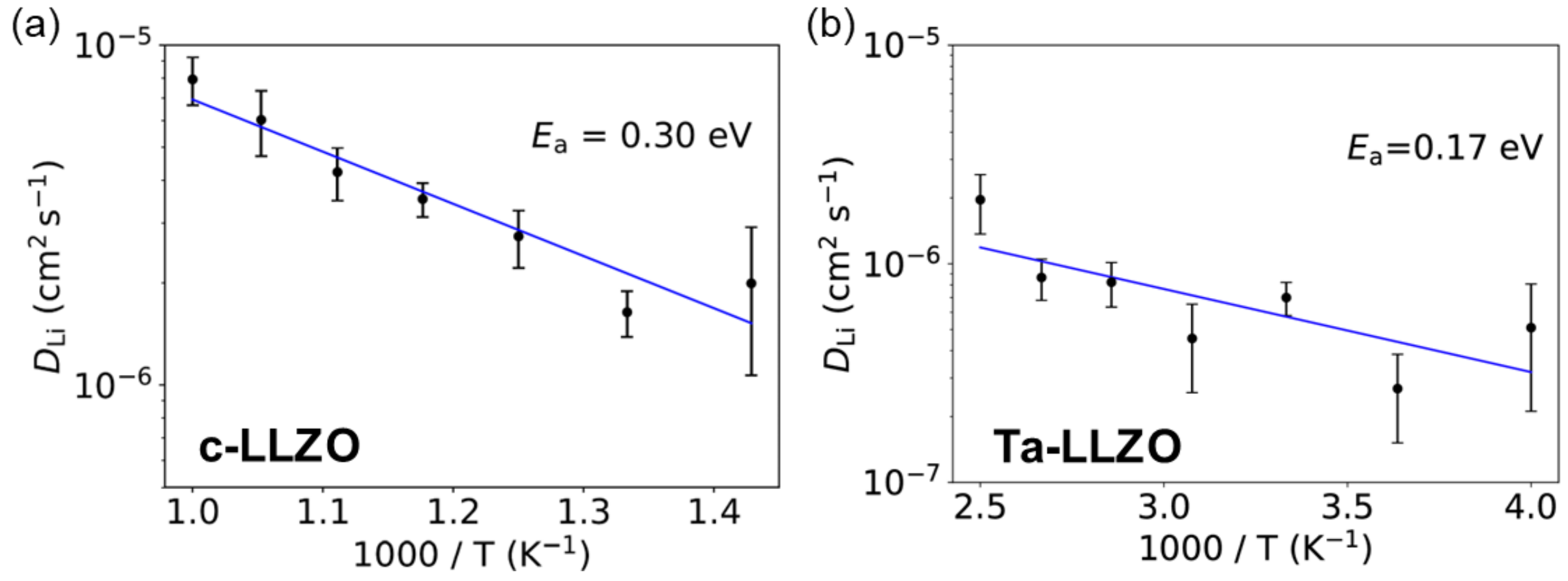

**Figure S5.** Arrhenius plots for (a) c-LLZO and (b) Ta-LLZO. At each temperature, four MLMD simulations of 200 ps were performed. The activation energies extracted from the slope of the diffusion coefficients are reported in each panel.

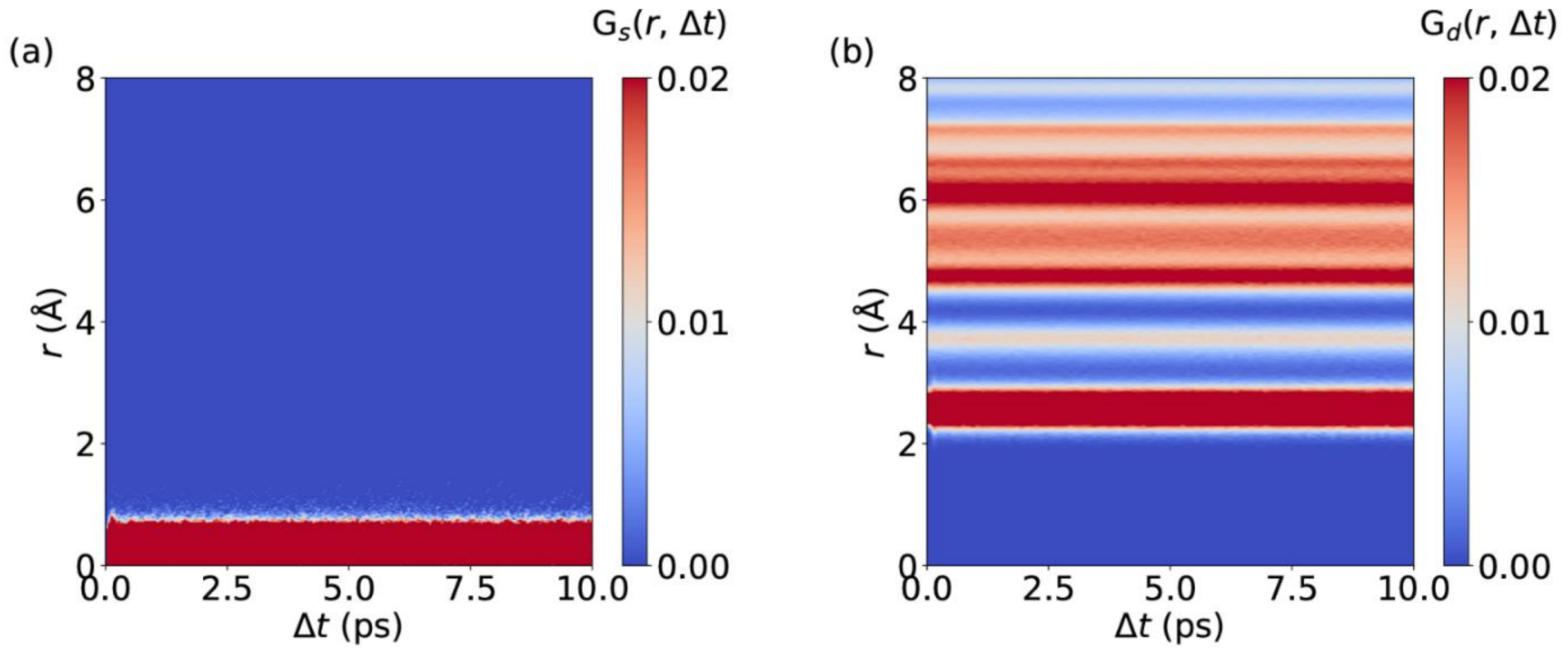

**Figure S6.** Spatiotemporal correlations of $Li^+$ ion in t-LLZO: (a) Self-part, $G_s(r, \Delta t)$, of the van Hove correlation function; (b) distinctive-part, $G_d(r, \Delta t)$, of the van Hove correlation functions. The absence of any changes in $G_d$ with time align with our observations of the absence of Li ion diffusion in t-LLZO.

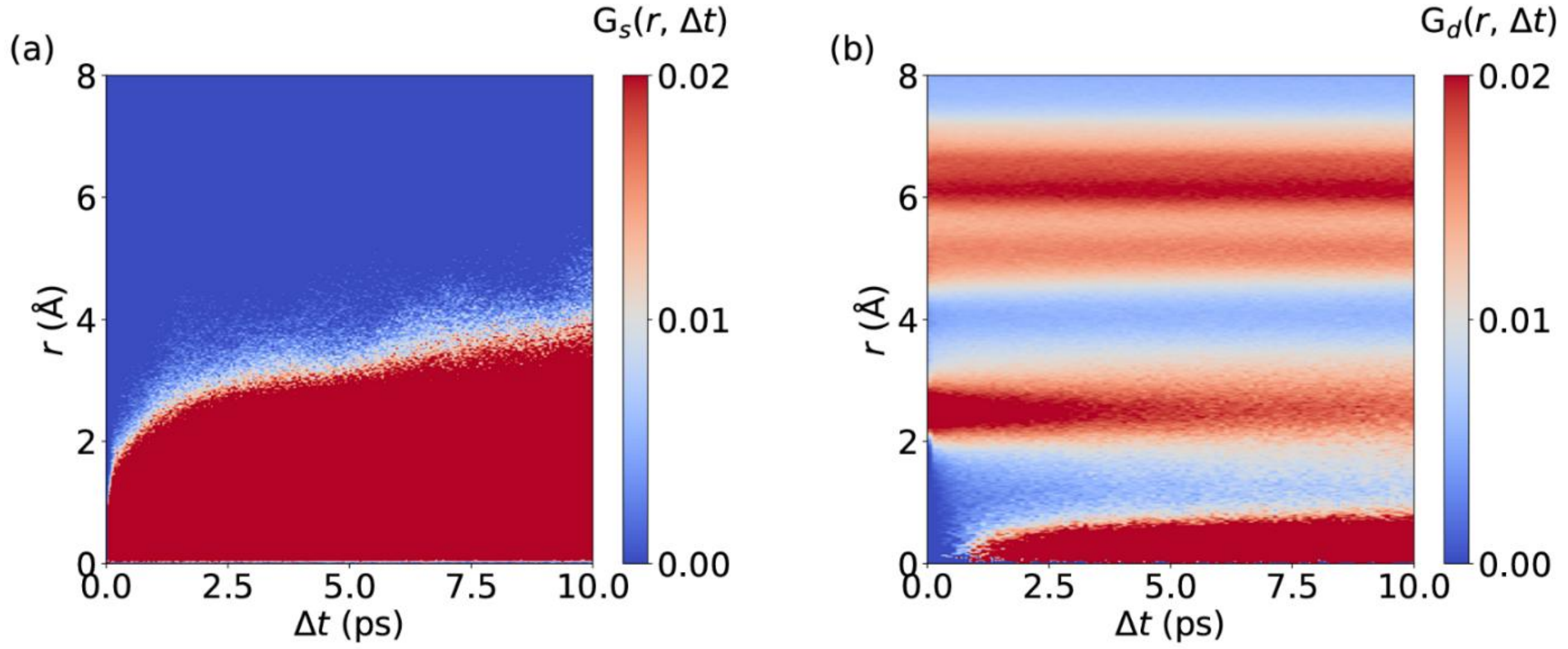

**Figure S7.** Spatiotemporal correlations of $Li^+$ ion in c-LLZO: (a) Self-part, $G_s(r, \Delta t)$, of the van Hove correlation function; (b) distinctive-part, $G_d(r, \Delta t)$, of the van Hove correlation functions. The strong changes in $G_d$ over time, with the branch at r ~ 0 Å arising within Δt of less than 2 ps, suggests the presence of concerted Li ion diffusion as demonstrated in previous work.[26,27]

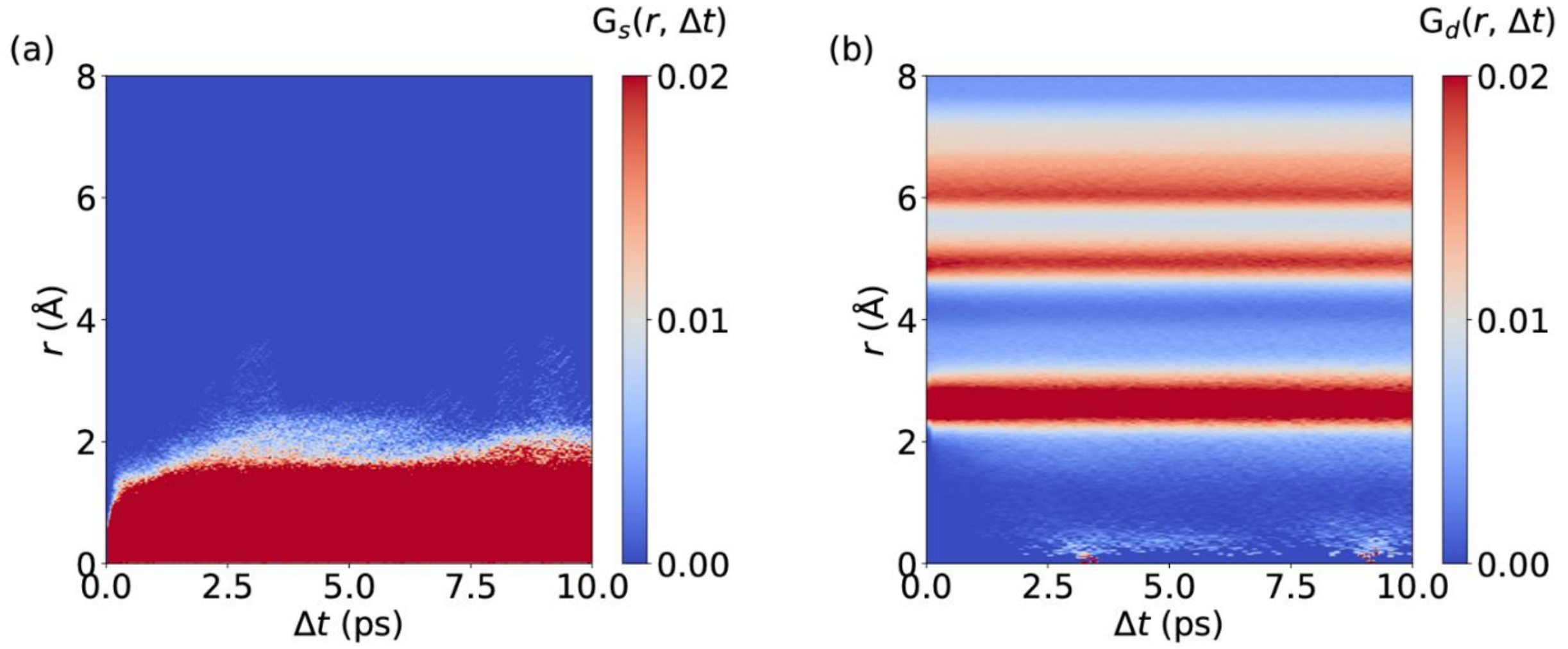

**Figure S8.** Spatiotemporal correlations of $Li^+$ ion in Ta-LLZO: (a) Self-part, $G_s(r, \Delta t)$, of the van Hove correlation function; (b) distinctive-part, $G_d(r, \Delta t)$, of the van Hove correlation functions. The small signatures visible in the $G_d$ at r ~ 0 Å and Δt of ~3 ps and ~ 8 ps, suggests the presence of Li ion transport via a thermally activated hopping mechanism.

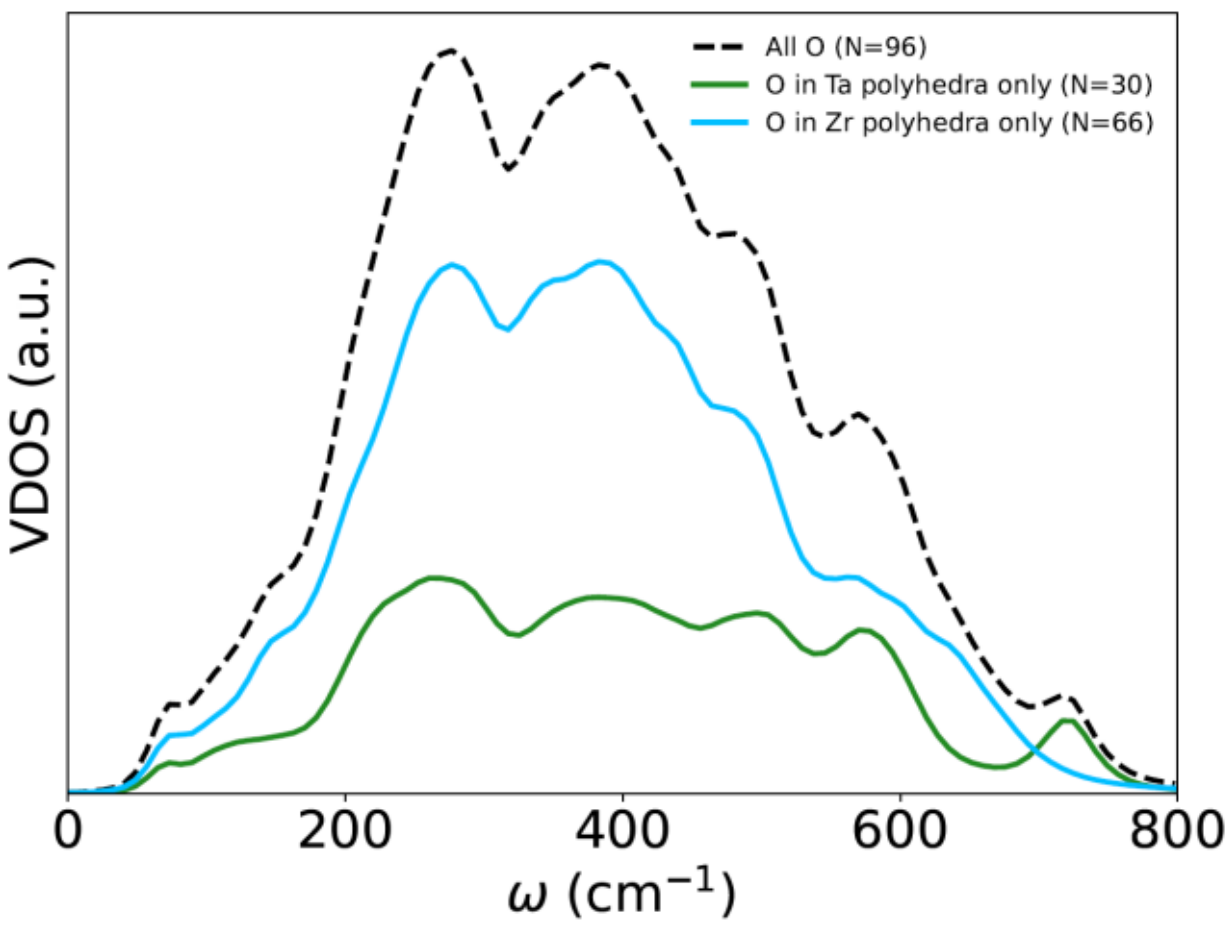

**Figure S9.** Contribution of oxygen atoms from $TaO_6$ and from $ZrO_6$ octahedra, as well as the total oxygen VDOS.

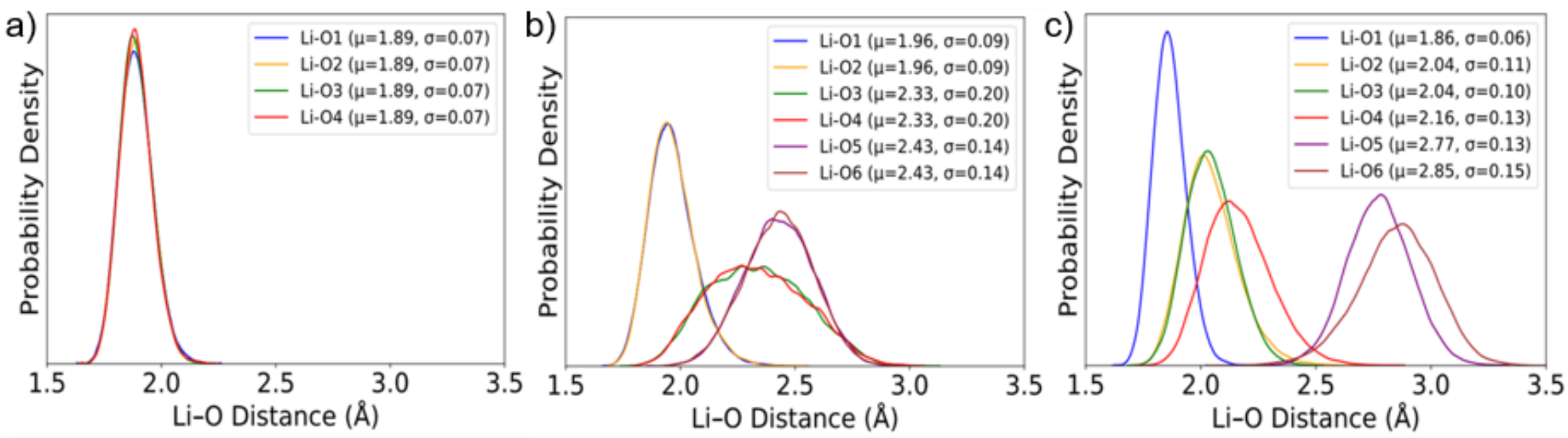

**Figure S10.** Distribution of Li-O bond lengths of the different lithium sites in t-LLZO. Panel a represents Li ions occupying a tetrahedral site, while panels b and c give Li-O bonds of Li ions in six-fold coordination with O ions. Average values and standard deviations of the Li-O bond distributions are specified for each Li-O bond.

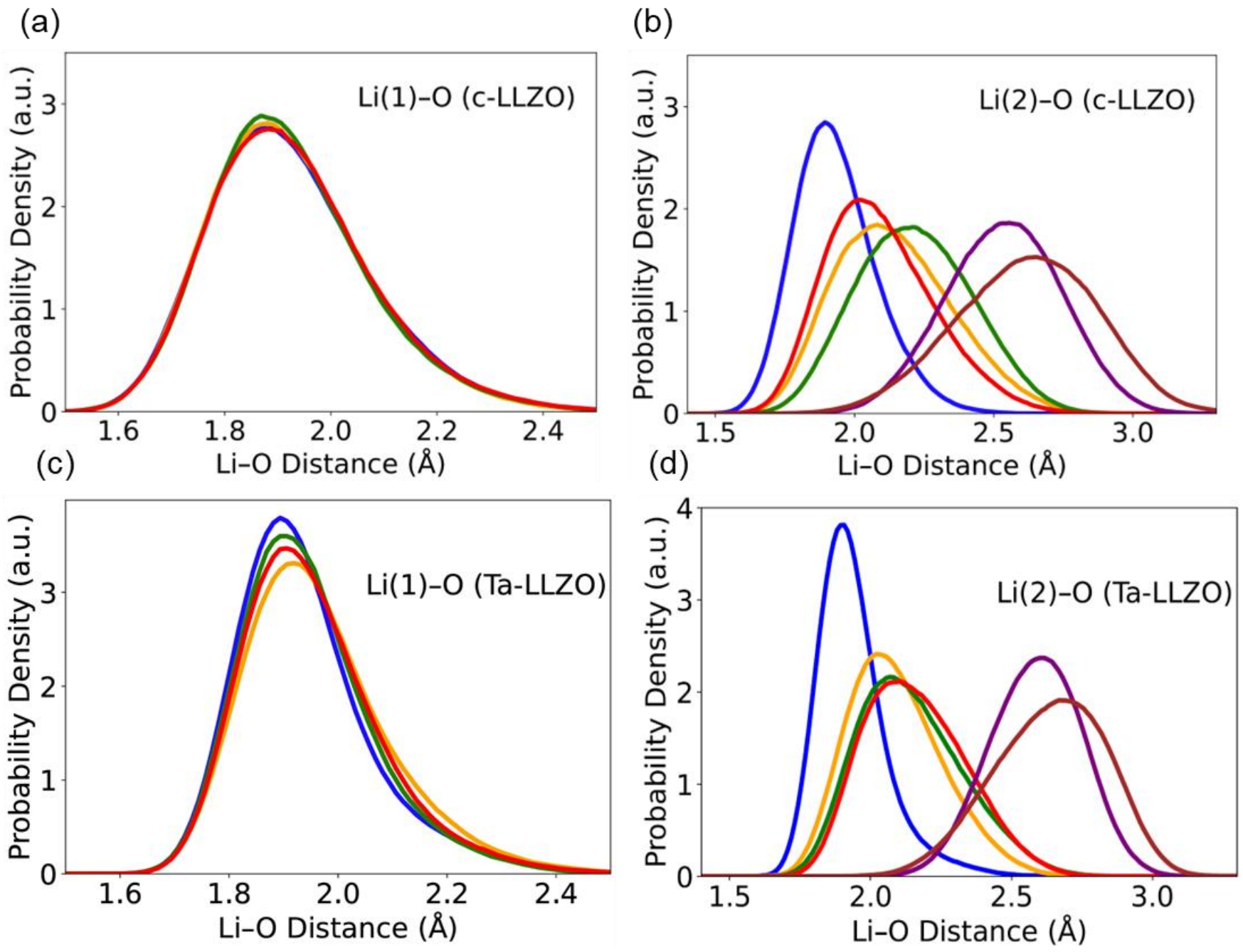

**Figure S11.** Distribution of Li-O bond lengths of the different lithium sites in (a, b) c-LLZO and (c, d) Ta-LLZO. Panels a and c visualize Li ions occupying a tetrahedral site, while panels b and d gives the Li-O bond distribution of Li ions in six-fold coordination with O ions.

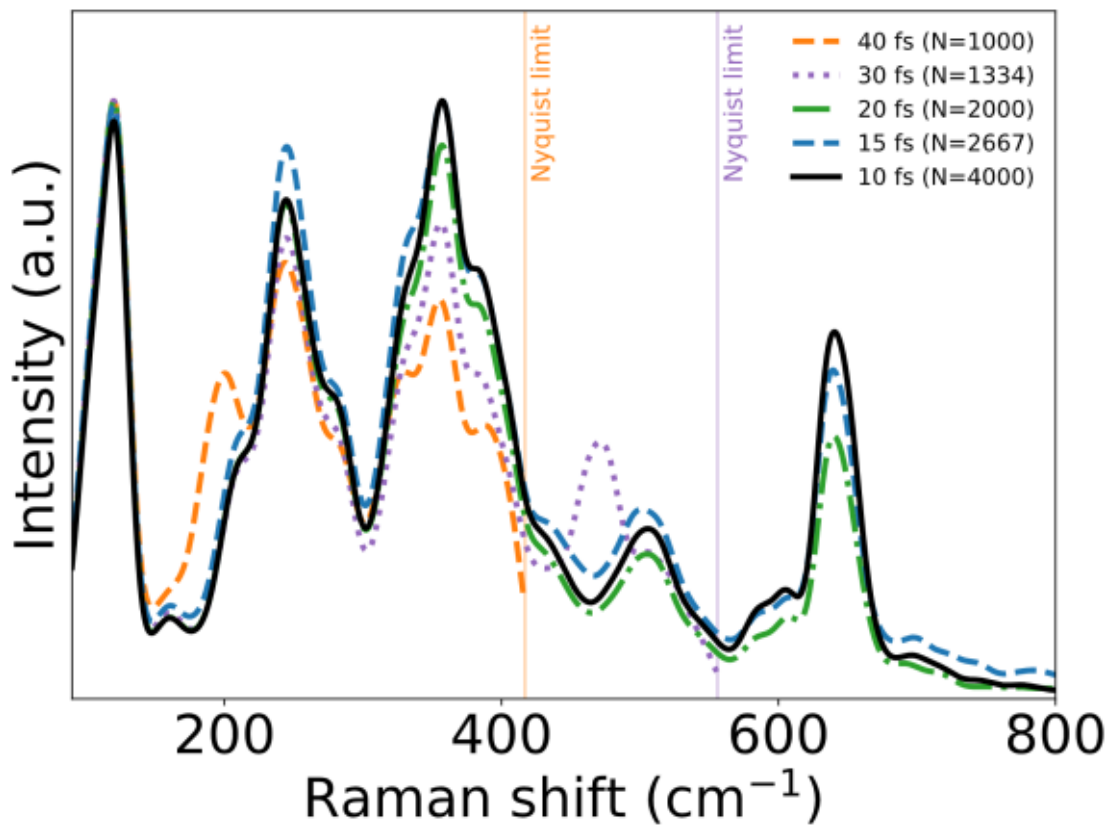

**Figure S12. Convergence of MD-Raman spectra with respect to the temporal spacing of DFPT calculations.** Raman spectra of t-LLZO using polarizability tensors evaluated every 10, 15, 20, 30, and 40 fs along the MLMD trajectory, corresponding to 40000, 2667, 2000, 1334, and 1000 DFPT calculations, respectively. Large sampling intervals lead to deviations in the high-frequency region due to the Nyquist limit. Production calculations employ a 10 fs sampling interval, ensuring convergence of all Raman features discussed in the main text.

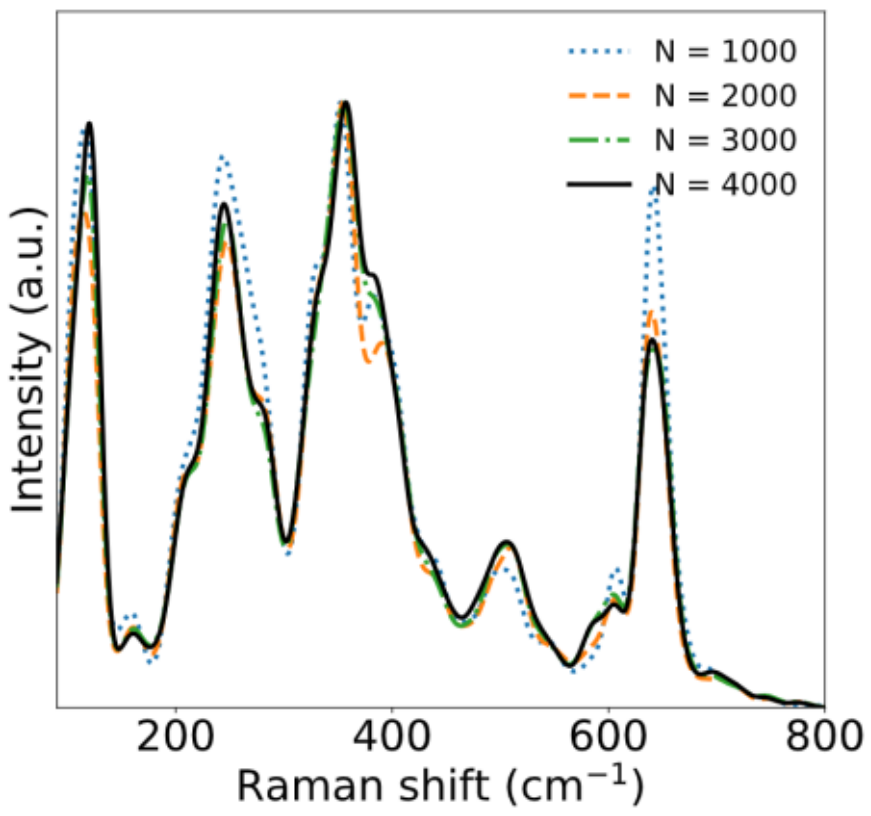

**Figure S13. Convergence of MD-Raman spectra with respect to trajectory length.** Raman spectra of t-LLZO for MD trajectory lengths of 10 ps (N = 1000), 20 ps (N = 2000), 30 ps (N = 3000), and 40 ps (N = 4000).

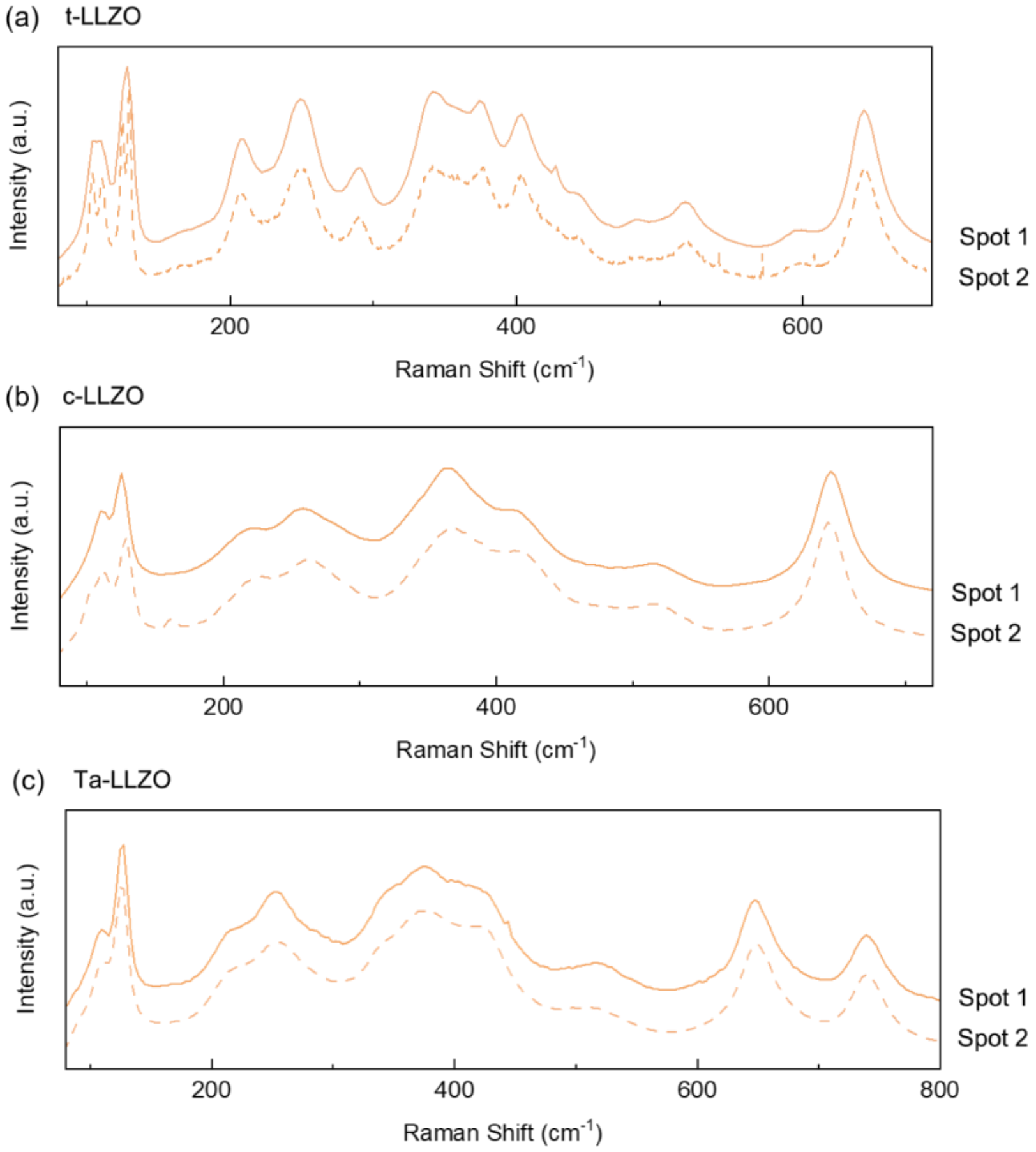

**Figure S14**. Spatial reproducibility of Raman spectra for tetragonal and cubic LLZO. Experimental Raman spectra of (a) tetragonal LLZO (t-LLZO), (b) Ga-doped cubic LLZO (c-LLZO), and (c) Ta-doped cubic LLZO (Ta-LLZO), collected at two different locations on each sample. Solid and dashed curves represent measurements from the two locations, respectively, and are vertically offset for clarity.

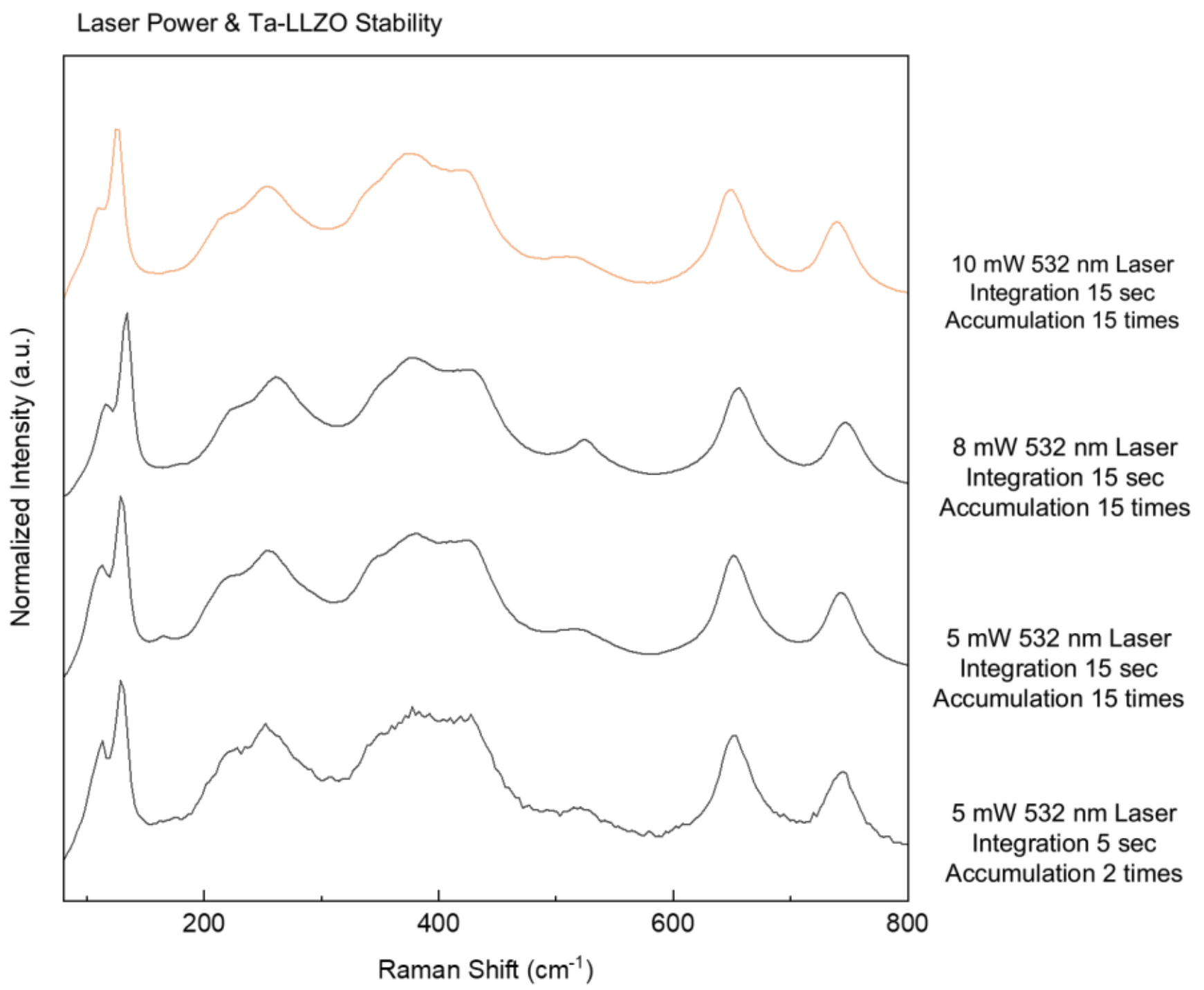

**Figure S15.** Influence of laser power and acquisition conditions on the Raman spectrum of Ta-LLZO. Raman spectra collected using a 532 nm excitation laser at powers of 10, 8, and 5 mW with an integration time of 15 s and 15 accumulations, and at 5 mW with an integration time of 5 s and 2 accumulations. The spectra are normalized and vertically offset for clarity.

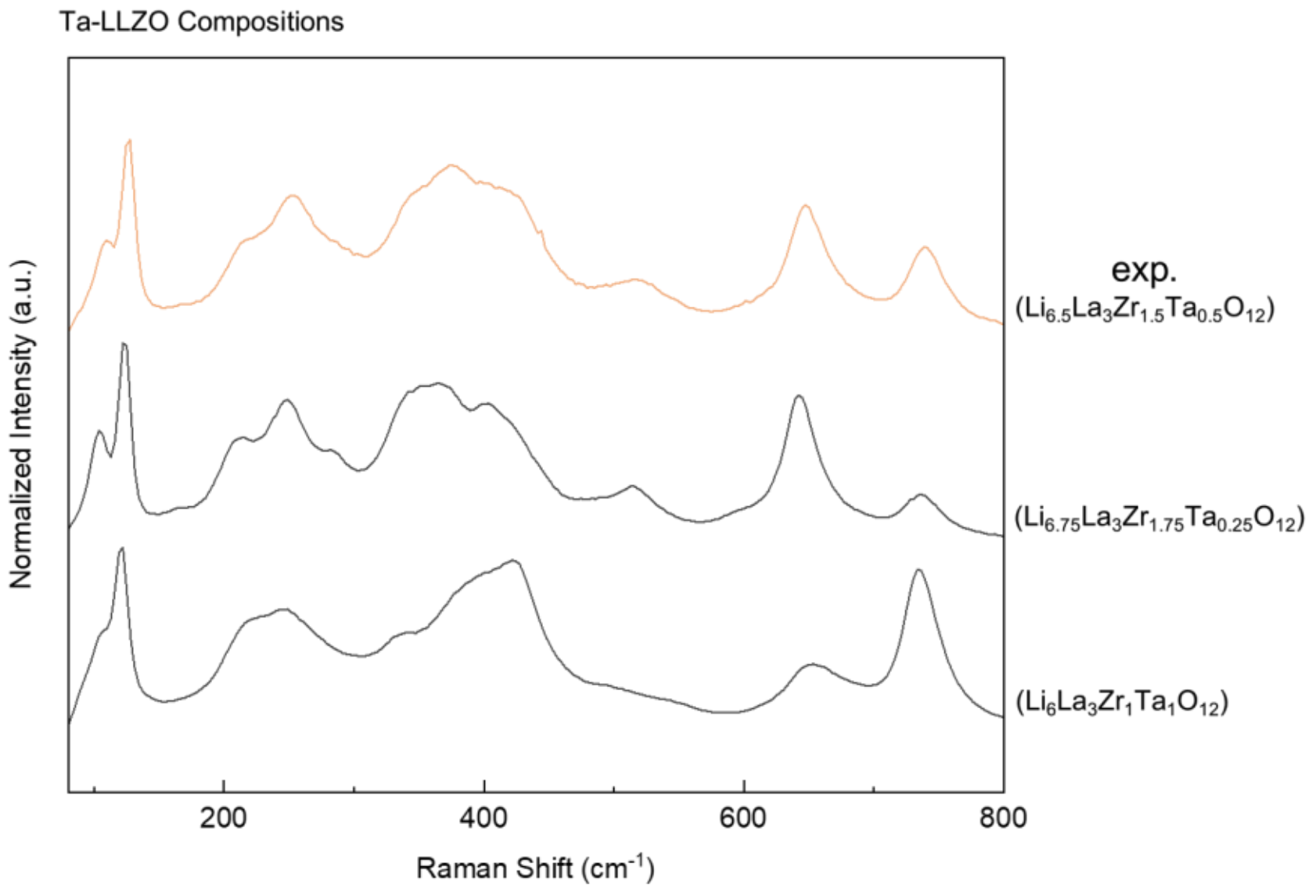

**Figure S16**. Raman spectra of Ta-doped LLZO with varying Ta concentrations. Experimental Raman spectra of $Li_{7-x}La_3Zr_{2-x}Ta_xO_{12}$ with x = 0.5, 0.25, and 1.0 (top to bottom). The spectra are normalized and vertically offset for clarity, highlighting changes in the Li-related vibrational region and the relative intensities of Zr–O and Ta–O stretching features with Ta substitution.

### Additional Tables

**Table S1.** Symmetry-resolved analysis of the Raman spectra of t-LLZO. Experimental and corresponding computed Raman shifts are reported, together with the percentage contribution of each symmetry channel to every feature. Most features draw on several channels rather than a single mode; the largest contribution is highlighted for readability only.

| **Raman shift ($cm^{-1}$)** | | **Symmetry Channels** | | | | |
|---|---|---|---|---|---|---|
| **Exp.** | **Comp.** | **$A_{1g}$** | **$B_{1g}$** | **$B_{2g}$** | **$E_g$** | **Dominant Channel** |
| 109 | 104 | 3.5 | 10.4 | 23.8 | 62.3 | $E_g$ |
| 128 | 121 | 47.7 | 47.8 | 1.3 | 3.2 | $B_{1g}$ |
| 171 | 161 | 13.9 | 5.9 | 10.7 | 69.4 | $E_g$ |
| 210 | 208 | 8.1 | 5.5 | 10.9 | 75.5 | $E_g$ |
| 250 | 243 | 49.5 | 32.0 | 11.9 | 6.7 | $A_{1g}$ |
| 291 | 282 | 36.0 | 31.6 | 3.8 | 38.6 | $A_{1g}$ |
| 345 | 331/358 | 81.0/27.9 | 10.7/38.0 | 1.2/1.2 | 7.1/32.9 | $A_{1g}$/$B_{1g}$ |
| 376 | 384 | 19.4 | 30.1 | 35.8 | 14.8 | $B_{2g}$ |
| 405 | 404 | 28.7 | 8.7 | 12.5 | 50.1 | $E_g$ |
| 429 | 427 | 21.2 | 6.4 | 32.3 | 40.2 | $E_g$ |
| 447 | - | - | - | - | - | - |
| 483 | 479 | 37.7 | 24.2 | 6.9 | 31.1 | $A_{1g}$ |
| 520 | 502/510 | 31.4/41.8 | 13.5/12.6 | 12.1/20.9 | 42.9/24.7 | $E_g$/$A_{1g}$ |
| 595 | 587/605 | 88.1/91.4 | 2.6/2.1 | 5.1/2.1 | 4.2/4.3 | $A_{1g}$/$A_{1g}$ |
| 644 | 638/648 | 98.1/98.5 | 0.4/0.3 | 0.5/0.5 | 1.0/0.7 | $A_{1g}$/$A_{1g}$ |

**Table S2.** Symmetry-resolved analysis of the Raman spectra of c-LLZO. Experimental and corresponding computed Raman shifts are reported, together with the percentage contribution of each symmetry channel to every feature. Most features draw on several channels rather than a single mode; the largest contribution is highlighted for readability only.

| **Raman shift ($cm^{-1}$)** | | **Symmetry Channels** | | | |
|---|---|---|---|---|---|
| **Exp.** | **Comp.** | **$A_{1g}$** | **$E_g$** | **$T_{2g}$** | **Dominant Channel** |
| 110 | 114 | 6.0 | 68.4 | 25.6 | $E_g$ |
| 126 | - | | | - | - |
| 218 | 193 | 12.9 | 29.9 | 57.2 | $T_{2g}$ |
| 260 | 225 | 11.0 | 49.2 | 39.8 | $E_g$ |
| 285 | 250 | 15.5 | 54.7 | 29.8 | $E_g$ |
| 365 | 342 | 22.0 | 59.5 | 18.5 | $E_g$ |
| 415 | 398 | 13.5 | 31.9 | 54.6 | $T_{2g}$ |
| 518 | 475/513 | 37.7/36.4 | 28.3/23.3 | 34.0/40.3 | $A_{1g}/A_{1g}$ |
| 580 | 548/580 | 56.4/74.8 | 18.9/9.9 | 24.7/15.3 | $A_{1g}/A_{1g}$ |
| 646 | 605/630/644 | 84.4/87.4/83.8 | 5.7/4.8/5.6 | 15.3/7.9/10.6 | $A_{1g}/A_{1g}/A_{1g}$ |

**Table S3.** Symmetry-resolved analysis of the Raman spectra of Ta-LLZO. Experimental and corresponding computed Raman shifts are reported, together with the percentage contribution of each symmetry channel to every feature. Most features draw on several channels rather than a single mode; the largest contribution is highlighted for readability only.

| **Raman shift ($cm^{-1}$)** | | **Symmetry Channels** | | | |
|---|---|---|---|---|---|
| **Exp.** | **Comp.** | **$A_{1g}$** | **$E_g$** | **$T_{2g}$** | **Dominant Channel** |
| 114 | 111 | 5.3 | 64.0 | 30.8 | $E_g$ |
| 133 | - | - | - | - | - |
| 220 | 218 | 8.5 | 51.9 | 39.6 | $E_g$ |
| 262 | 250 | 9.2 | 67.6 | 23.1 | $E_g$ |
| 353 | 315 | 17.6 | 65.4 | 17.0 | $E_g$ |
| 377 | 356 | 14.0 | 57.9 | 28.1 | $E_g$ |
| 430 | 394 | 12.5 | 24.2 | 63.2 | $T_{2g}$ |
| 526 | 495 | 14.0 | 37.5 | 48.5 | $T_{2g}$ |
| 655 | 635 | 91.9 | 3.2 | 5.0 | $A_{1g}$ |
| 747 | 705/725 | 93.9/95.1 | 2.1/1.9 | 4.0/3.1 | $A_{1g}$/$A_{1g}$ |

**Table S4.** Summary of LLZO compositions, Li-ion transport properties, and experimental Raman measurement conditions.

| Parameter | t-LLZO | c-LLZO | Ta-LLZO |
|---|---|---|---|
| Computational composition | $Li_7La_3Zr_2O_{12}$ | $Li_7La_3Zr_2O_{12}$ | $Li_{6.375}La_3Zr_{1.375}Ta_{0.625}O_{12}$ |
| Experimental composition | $Li_7La_3Zr_2O_{12}$ | $Li_{6.4}Ga_{0.2}La_3Zr_2O_{12}$ | $Li_{6.5}La_3Zr_{1.5}Ta_{0.5}O_{12}$ |
| MLMD temperature (K) | 300 | 900 | 300 |
| Li diffusion coefficient ($cm^2$ $s^{-1}$) | Not resolved | ~$4.2 \times 10^{-6}$ | ~$7.0 \times 10^{-7}$ |
| Activation energy (eV) | N/A | 0.30 | 0.17 |
| Experimental Raman temperature | RT | RT | RT |
| Laser wavelength (nm) | 532 | 532 | 532 |
| Laser power (mW) | 10 | 10 | 10 |
| Integration time (s) | 15 | 15 | 15 |
| Accumulations | 15 | 15 | 15 |